\newcommand{\be}{\begin{equation}}
\newcommand{\ee}{\end{equation}}
\newcommand{\brr}{\begin{eqnarray}}
\newcommand{\err}{\end{eqnarray}}
\newcommand{\nn}{\nonumber}
\newcommand{\bd}{\begin{displaymath}}
\newcommand{\ed}{\end{displaymath}}

\newcommand{\bib}{\bibitem}
\newcommand{\bfig}{\begin{figure}}
\newcommand{\efig}{\end{figure}}
\newcommand{\ie}{i.e.}
\newcommand{\eg}{e.g.}
\DeclareMathAlphabet{\mathpzc}{OT1}{pzc}{m}{it}
\def\alf{\alpha}
\def\bet{\beta}
\def\gam{\gamma}
\def\th{\theta}

\def\om{\omega}
\def\eps{\varepsilon}
\def\rpar{\right)}
\def\lpar{\left(}
\def\rbk{\right]}
\def\lbk{\left[}
\def\rbr{\right\}}
\def\lbr{\left\{}
\def\lb{\label}

\def\bop{\mbox{\boldmath $\mathrm{B}$}}
\def\nop{\mbox{\boldmath $\mathrm{N}$}}
\def\opexp{\mbox{$\bcal{E}$}}
\def\opcos{\mbox{$\bcal{C}$}}
\def\opsin{\mbox{$\bcal{S}$}}
\def\indq{\mbox{\tiny $q$}}
\def\inrs{\mbox{\tiny RS}}
\def\inho{\mbox{\tiny HO}}
\def\incs{\mbox{\tiny $\mathcal{CS}$}}
\def\inc{\mbox{\tiny $\mathcal{C}$}}
\def\ins{\mbox{\tiny $\mathcal{S}$}}
\def\inn{\mbox{\tiny $\mathrm{N}$}}
\def\rg{\rangle}
\def\lg{\langle}
\def\nc{\mathrm{i}}
\def\prz{\mathscr{H}}
\def\arz{\mathscr{A}}

\def\erz{\mathscr{E}}

\def\pq{\mathfrak{q}}
\def\coloneq{\mathrel{\mathop:}=}
\def\half{\case{1}{2}}

\documentclass{iopart}
\usepackage{iopams}
\usepackage{mathrsfs}
\usepackage{upmath}
\usepackage{yfonts}
\usepackage{dsfont}
\usepackage{bbm}
\usepackage{graphicx}
\usepackage[all]{xy}
\usepackage{ifsym}
\usepackage{pifont}
\begin{document}
%%%%%%%%%%%%%%%%%%%%%%%%%%%%%%%%%%%%%%%%%%%%%%%%%%%%%%%%%%%%%%%%%%%%%%%%%%%%%%%%%%%%%%%%%%%%%%%%%%%%%%%%%%%%%%%%%%%%%%%%%%%%%%%%%%%%%%%%
\title[{\rm M A Marchiolli} et al]{Algebraic properties of Rogers-Szeg\"{o} functions: I. Applications in quantum optics}
\author{M A Marchiolli, M Ruzzi and D Galetti}
\address{Instituto de F\'{\i}sica Te\'{o}rica, Universidade Estadual Paulista,
         Rua Pamplona 145, 01405-900 S\~{a}o Paulo, SP, Brazil}
\ead{mamarchi@ift.unesp.br}
%%%%%%%%%%%%%%%%%%%%%%%%%%%%%%%%%%%%%%%%%%%%%%%%%%%%%%%%%%%%%%%%%%%%%%%%%%%%%%%%%%%%%%%%%%%%%%%%%%%%%%%%%%%%%%%%%%%%%%%%%%%%%%%%%%%%%%%%
\begin{abstract}
By means of a well-established algebraic framework, Rogers-Szeg\"{o} functions associated with a circular geometry in the complex plane
are introduced in the context of $q$-special functions, and their properties are discussed in details. The eigenfunctions related to the
coherent and phase states emerge from this formalism as infinite expansions of Rogers-Szeg\"{o} functions, the coefficients being 
determined through proper eigenvalue equations in each situation. Furthermore, a complementary study on the Robertson-Schr\"{o}dinger and 
symmetrical uncertainty relations for the cosine, sine and nondeformed number operators is also conducted, corroborating, in this way,
certain features of $q$-deformed coherent states. 
\end{abstract}
\submitto{{\it J. Phys. A: Math. Theor.}}
%%%%%%%%%%%%%%%%%%%%%%%%%%%%%%%%%%%%%%%%%%%%%%%%%%%%%%%%%%%%%%%%%%%%%%%%%%%%%%%%%%%%%%%%%%%%%%%%%%%%%%%%%%%%%%%%%%%%%%%%%%%%%%%%%%%%%%%%
\section{Introduction}
%%%%%%%%%%%%%%%%%%%%%%%%%%%%%%%%%%%%%%%%%%%%%%%%%%%%%%%%%%%%%%%%%%%%%%%%%%%%%%%%%%%%%%%%%%%%%%%%%%%%%%%%%%%%%%%%%%%%%%%%%%%%%%%%%%%%%%%%

The Ramanujan's work and subsequent studies on the $q$-special functions certainly represent an important chapter within several 
astonishing achievements reached along the last decades in mathematics \cite{Askey0,Ramanujan,Askey1,Askey2,Ismail,Gelfand,Andrews,Exton,
Gasper,Vilenkin}. Another particular but not least remarkable research branch developed in parallel by mathematicians and physicists is
focussed on quantum groups and/or $q$-deformed algebras \cite{Chari,Lohe}. Since $q$-deformed algebras encompass the description of a wide 
variety of symmetries, if one compares with those studied in the standard Lie algebras, it turned natural to employ this powerful mathematical 
tool to investigate certain complex symmetries associated with nontrivial physical systems in an appropriate way. Hence, several contributions 
have appeared in the literature, from time to time, presenting a plethora of original results directly related to specific problems 
originated from solvable statistical mechanics models \cite{Baxter}, quantum inverse scattering theory \cite{Korepin}, nuclear physics
\cite{Bonatsos,Bahri}, molecular physics \cite{Kotsos}, some $q$-deformed extensions of quantum mechanics \cite{Macfarlane,Gould,Burban,
Rubin,Lavagno} and quantum optics with emphasis on coherent states \cite{Arik,Gray,Nelson,Perelomov,Borzov,Quesne,Marchiolli,Ngompe},
as well as recent applications in trapped ions by laser fields \cite{Vogel} and also in the Jaynes-Cummings model \cite{Geloun}. Furthermore, 
it is worth mentioning that this range of possible applications can also be extended to deformed superalgebras \cite{Chaichian}, knot theories
\cite{Kauffman} and non-commutative geometries \cite{Manin}.

In this prominent scenario, there are approaches that connect $q$-special functions and $q$-deformed algebras which deserve be placed in
evidence because they exhibit a unifying quantum-algebraic framework for the realizations and/or representations of those algebras. For example, 
within the context of Lie algebras and their $q$-analogues, Feinsilver \cite{Feinsilver} has discussed how the three-term recurrence relations 
related to orthogonal polynomials can be used to obtain certain realizations in terms of raising and lowering operators. Floreanini and Vinet
\cite{Floreanini} showed that suitable operators acting on vector spaces of functions in one complex variable can be considered as possible
realizations of the $\mathfrak{sl}_{q}(2)$ and $q$-oscillator algebras. Pursuing this formal line of theoretical investigation, some few authors
\cite{Atakishiyev,Rahman,Galetti} also presented significant contributions for a special class of orthogonal polynomials, which are recognized 
in the literature as Rogers-Szeg\"{o} (RS) and Stieltjes-Wigert (SW) polynomials \cite{Rogers,Szego,Carlitz}. The mathematical motivation for this
specific choice of polynomials is directly associated with the $q$-oscillator algebra, namely, both the polynomials are viewed as concrete
representations of the Iwata-Arik-Coon-Kuryshkin (IACK) algebra \cite{Burban}, and their respective realizations expressed by means of the Jackson's
$q$-derivative \cite{Jackson1,Jackson2}. In addition, while the SW polynomials are orthogonalized on the full real line, the RS polynomials are 
defined upon the complex plane and orthogonalized on the unit circle through a particular measure \cite{Galetti,Carlitz}. This fundamental feature
intrinsic to the RS polynomials has a potential connection with angular representations in quantum mechanics, such fact being interpreted by us as 
an effective gain in what concerns the debate on the polar decomposition of the annihilation operator in quantum optics
\cite{Susskind,Carruthers,Bergou,Review,Perina}.

The main goal of this paper is to present a consistent algebraic framework based on a particular set of $q$-special functions which leads us 
to go further in our comprehension on the phase operator problem and its different representations in quantum mechanics. For this purpose,
we first review certain essential mathematical properties associated with the RS polynomials which permit us to establish two new (as far as
we know) integral representations for such polynomials involving the finite $q$-Pochhammer symbol and the SW polynomials. In the following, we
define our object of study (here named as RS functions) through a product of two complex functions with distinct essential features, namely
$\Psi_{n}^{\inrs}(z;q) \coloneq \mathscr{R}_{n}(z;q) \mathscr{M}(z;q)$. Indeed, while the first one is a RS polynomial (at least of the 
normalization coefficient), the second one is responsible for the corresponding weight function which is connected with the decomposition of the
Szeg\"{o} measure in the complex plane. An immediate consequence from this peculiar sort of definition refers to its inherent orthogonality property,
that is, it preserves the orthogonality relation verified for the RS polynomials. Another important property concerns the completeness relation for 
the RS functions that is presented in this context via a bilinear kernel \cite{Atakishiyev}. The $q$-calculus framework is then employed to carry out 
a careful analysis of each aforementioned complex function, and the results originated from this analysis used to derive the $q$-differential forms 
of the lowering, raising and number operators. It is worth stressing that the algebraic approach here developed for the RS functions leads us to
obtain, in principle, not only an alternative representation for the IACK algebra but also an inherent realization. 

The second part of this paper is focussed basically on the construction process of coherent and phase states in accordance with the 
quantum-algebraic framework previously discussed. So, our first application has as reference guide the mathematical approach developed in 
\cite{Barut} for an important class of coherent states, namely, those obtained from a determined eigenvalue equation for a given annihilation 
operator \cite{Arik,Gray,Nelson,Perelomov,Borzov,Quesne,Marchiolli,Ngompe,Vogel}. The eigenfunctions derived from this particular procedure are 
then expressed as an infinite expansion in terms of the RS functions whose coefficients satisfy a set of mathematical prerequesites that leads 
us to obtain, as a by-product, the excitation probability distribution for the $q$-deformed coherent state. Expressing in a clearer way, once 
the physical system of interest can be initially prepared in the $q$-deformed coherent state, this result allows us to obtain the excitation
probability distribution (here labelled by the degree of excitation $n$ of the $q$-deformed harmonic oscillator) for such system. Furthermore, 
we discuss in details some intrinsic properties of this definition. In what concerns the phase states and their connections with the angular
representations in quantum mechanics, we have constructed the $q$-deformed eigenstates for the cosine and sine operators following the 
Carruthers-Nieto approach \cite{Carruthers}, and also presented the orthogonality and completeness relations for each situation. To complete this 
work, we have applied our results in order to calculate, via $q$-deformed coherent states, certain mean values associated with the cosine and sine
operators which allow us to carry out a detailed study on the Robertson-Schr\"{o}dinger and symmetrical uncertainty relations. 

This paper is structured as follows. In section 2 we establish some few essential mathematical properties and also derive two additional integral
representations for the RS polynomials. In section 3 we introduce the RS functions $\{ \Psi_{n}^{\inrs}(z;q) \}_{n \in \mathbb{N}}$ through the 
product of two basic complex functions whose distinct characteristics lead us to obtain a wide set of results which constitutes our quantum-algebraic
framework. In the following, section 4 is dedicated to the construction process of $q$-deformed coherent states related to the RS functions, where
certain properties (for instance, the overlap probability and completeness relation) are discussed in details. Moreover, we obtain in section 5 the
respective eigenstates of the cosine and sine operators, as well as present some relevant results for each situation. The discussion on the
Robertson-Schr\"{o}dinger and symmetrical uncertainty relations involving such operators is presented in section 6. Finally, section 7 contains our
summary and conclusions.     

%%%%%%%%%%%%%%%%%%%%%%%%%%%%%%%%%%%%%%%%%%%%%%%%%%%%%%%%%%%%%%%%%%%%%%%%%%%%%%%%%%%%%%%%%%%%%%%%%%%%%%%%%%%%%%%%%%%%%%%%%%%%%%%%%%%%%%%%
\section{Explanatory notes on the RS polynomials}
%%%%%%%%%%%%%%%%%%%%%%%%%%%%%%%%%%%%%%%%%%%%%%%%%%%%%%%%%%%%%%%%%%%%%%%%%%%%%%%%%%%%%%%%%%%%%%%%%%%%%%%%%%%%%%%%%%%%%%%%%%%%%%%%%%%%%%%%

In order to make the presentation of this section more self-contained, let us initially review certain essential mathematical prerequisites 
of the RS polynomials, for then establishing, subsequently, two integral representations which permit to connect such polynomials with
the finite $q$-Pochhammer symbol and the SW polynomials. It is important to emphasize that the basic notation employed in our exposition
follows some well-known textbooks on special functions, where, in particular, the $q$-series or Eulerian series are introduced within the 
context of theory of partitions and/or basic hypergeometric series \cite{Andrews,Exton,Gasper,Vilenkin}.

\subsubsection*{{\bf Definition.}}
{\it Let $\{ \prz_{n}(z;q) \}_{n \in \mathbb{N}}$ designate a set of polynomials with $0 < q < 1$ and $z \in \mathbb{C}$. In particular, 
the RS polynomials are defined through of the finite series} \cite{Rogers,Szego} {\it
\be
\lb{e1}
\quad \prz_{n}(z;q) \coloneq \sum_{k=0}^{n} \lbk \begin{array}{c} n \\ k \\ \end{array} \rbk_{\indq} z^{k} ,
\ee
where the $q$-binomial coefficients (also known as Gaussian polynomials)
\bd
\lbk \begin{array}{c} n \\ k \\ \end{array} \rbk_{\indq} \coloneq \frac{(q;q)_{n}}{(q;q)_{k} (q;q)_{n-k}} = 
\frac{[ n ]_{\indq}!}{[ k ]_{\indq}! [ n-k ]_{\indq}!}
\ed
are expressed in terms of the finite $q$-Pochhammer symbol
\bd
(a;q)_{n} \coloneq \prod_{j=0}^{n-1} (1 - a q^{j}) \qquad (a \in \mathbb{C})
\ed
or written as a function of the $q$-factorial $[ n ]_{\indq}! \coloneq ( 1 - q )^{-n} (q;q)_{n}$.}

It is worth noticing that the generating function associated with $\prz_{n}(z;q)$ is given by the specific Eulerian series $G(w,z;q) 
\coloneq \lbk (w;q)_{\infty} (wz;q)_{\infty} \rbk^{-1}$ for $| wz | < 1$. Thus, for future use in the text, in order to establish a consistent 
proof of this statement, let us first consider the particular case $G(w,0;q) = \lbk (w;q)_{\infty} \rbk^{-1}$. In this situation, the Cauchy 
theorem states that, for $| w | < 1$ and $0 < q < 1$, $G(w,0;q)$ is reduced to the infinite series \cite{Andrews}
\bd
G(w,0;q) = \sum_{n \in \mathbb{N}} \frac{w^{n}}{(q;q)_{n}} \qquad (w \in \mathbb{C}) .
\ed
This result leads us to propose that $G(w,z;q)$ admits a similar expression, where now the summand is multiplied by the coefficient
$\arz_{n}(z;q)$,
\be
\lb{e2}
G(w,z;q) = \sum_{n \in \mathbb{N}} \arz_{n}(z;q) \frac{w^{n}}{(q;q)_{n}} .
\ee
Indeed, equation (\ref{e2}) has the essential features that we need for our purposes since the infinite product $G(w,z;q)$ is uniformly 
convergent for all $q$ inside $| wz | \leq 1 - \eps$, and therefore it defines a function of $w$ and $z$ analytic in $| wz | < 1$. 
Furthermore, the identity \cite{Carlitz}
\bd
G(qw,z;q) = (1-w) (1-qw) G(w,z;q)
\ed
allows us to show that $\{ \arz_{n}(z;q) \}_{n \in \mathbb{N}}$ satisfies a three-term recurrence relation
\be
\lb{e3}
\arz_{n+1}(z;q) = (1+z) \arz_{n}(z;q) - (1 - q^{n}) z \arz_{n-1}(z;q) ,
\ee
where $\arz_{0}(z;q) = 1$ and $\arz_{1}(z;q) = 1 + z$. The remaining terms for $n \geq 2$ determine a set of polynomials expressed
explicitly in terms of $z$ and $q$, whose closed formula coincides exactly with equation (\ref{e1}); consequently, $\arz_{n}(z;q) 
\equiv \prz_{n}(z;q)$. \ding{114}

Next, let us introduce some few properties related to the RS polynomials where special attention will be paid to their orthogonality 
property. Adopting a particular parametrization for the complex variable $z$, Szeg\"{o} \cite{Szego} showed that $\{ \prz_{n}(z;q) 
\}_{n \in \mathbb{N}}$ can be orthogonalized on the circle through a specific measure which coincides with the Jacobi 
$\vartheta_{3}$-function evaluated at continuous arguments \cite{WW}, that is,
\be
\lb{e4}
\fl \quad \int_{- \pi}^{\pi} \prz_{m} \lpar - q^{-\half} e^{- \nc \varphi}; q \rpar \prz_{n} \lpar - q^{-\half} e^{\nc \varphi}; q 
\rpar \vartheta_{3} \lpar \frac{\varphi}{2} \biggl| q^{\half} \rpar \frac{d \varphi}{2 \pi} = \frac{(q;q)_{n}}{q^{n}} \, 
\delta_{m,n} .
\ee
Subsequently, Carlitz \cite{Carlitz} generalized such equation by fixing the integration measure and considering different arguments of
$\prz_{n}(z;q)$. Since then, different authors have worked on this theme and showed some interesting peculiarities of the RS polynomials. 
For example, Macfarlane \cite{Macfarlane} has discussed the quantum group ${\rm SU}_{q}(2)$ through a mathematical procedure that resembles 
the approach developed by Schwinger \cite{Schwinger} for the quantum theory of angular momentum. In particular, the author showed how the 
coordinate representation of the $q$-deformed harmonic oscillator can be used in order to obtain a wavefunction which is expressed in terms 
of RS polynomials. Moreover, Atakishiyev and Nagiyev \cite{Atakishiyev} derived an important orthogonality relation on the full real line 
for such polynomials, and also established a special link with the SW polynomials by means of a Fourier transform. Recently, Galetti and 
coworkers \cite{Galetti} have shown didactically both the orthogonality relations for the RS and SW polynomials, as well as obtained the 
explicit realizations of the raising and lowering operators for each case;\footnote{It is worth mentioning that Ismail and Rahman \cite{Rahman}
slightly modified the argument of $\prz_{n}(z;q)$ (namely, $z \rightarrow q^{- 1/2} z$) with the aim of deriving the respective 
$q$-differential forms related to the raising and lowering operators. This particular procedure has then produced certain ladder operators for 
the RS polynomials whose formal expressions differ from those obtained in \cite{Galetti}.} in addition, the authors also proposed a Wigner 
function related to the RS polynomials which leads us to determine a set of well-behaved marginal distribution functions with compact support 
for the angle and action variables.

The next property to be discussed allows us to establish a connection between the finite $q$-Pochhammer symbol and the RS polynomials.
For this task let us initially express, by means of the Cauchy theorem, $(a;q)_{n}$ as a sum involving finite powers of the complex
variable $a$, namely
\bd
(a;q)_{n} = \sum_{k=0}^{n} \lbk \begin{array}{c} n \\ k \\ \end{array} \rbk_{\indq} q^{\half k(k-1)} (- a)^{k} .
\ed
This result is extremely important in our considerations since it leads us to verify that
\be
\lb{e5}
\int_{- \pi}^{\pi} \prz_{n} \lpar - q^{-\half} a^{r} e^{\nc s \varphi}; q \rpar \vartheta_{3} \lpar u \varphi \biggl| 
q^{\case{2 u^{2}}{s^{2}}} \rpar \frac{d \varphi}{2 \pi} = \lpar a^{r};q \rpar_{n} , 
\ee
where $r$ is an arbitrary power of $a$ and $\case{s}{2u}$ is an integer number --- note that the right-hand side of equation (\ref{e5})
does not depend on $s$ and $u$ in this situation. Thus, if one substitutes $r=1$ in such case, this identity can be considered as a possible 
integral representation for the finite $q$-Pochhammer symbol. Another interesting additional property consists in establishing an integral 
representation for the RS polynomials through an adequate transformation kernel $\mathpzc{P}_{r}(\om;q)$, that is,
\be
\lb{e6}
\int_{0}^{\infty} \lpar - q^{r} z \om; q \rpar_{n} \mathpzc{P}_{r}(\om;q) \, d \om = \prz_{n}(z;q)
\ee
with
\bd
\fl \qquad \mathpzc{P}_{r}(\om;q) \coloneq \frac{m}{\sqrt{\pi}} \, \om^{r - \case{3}{2}} \exp \lbk - \frac{1}{4 m^{2}} \lpar r - 
\half \rpar^{2} - m^{2} \ln^{2} (\om) \rbk \qquad (m \in \mathbb{R})
\ed
and $q = \exp \lpar - \case{1}{2 m^{2}} \rpar$. In order to demonstrate such equation, it is sufficient to know that
\be
\lb{e7}
\int_{0}^{\infty} \om^{k} \mathpzc{P}_{r}(\om;q) \, d \om = q^{- \half k(k-1) - rk}
\ee
for the following proper parametrization: $\om = \exp \lpar \case{x}{m^{2}} \rpar$ $(- \infty < x < \infty)$. It is worth noticing 
that $\mathpzc{P}_{r}(\om;q)$ not only encompasses certain particular cases studied by Carlitz \cite{Carlitz}, but also can be used 
to orthogonalize the SW polynomials.

Let us derive now a last integral representation for the RS polynomials which has, as an integrand, the product of the SW polynomials
\be
\lb{e8}
\mathscr{G}_{n}(q^{n+r} z \om;q) \coloneq \sum_{k=0}^{n} \lbk \begin{array}{c} n \\ k \\ \end{array} \rbk_{\indq} q^{k(k-n)} 
\lpar q^{n+r} z \om \rpar^{k}
\ee
and the transformation kernel
\bd
\mathpzc{Y}_{r}(\om;q) \coloneq \frac{m}{\sqrt{2 \pi}} \, \om^{\case{r}{2}-1} \exp \lbk - \frac{r^{2}}{8 m^{2}} - \frac{m^{2}}{2}
\ln^{2} (\om) \rbk .
\ed
In this situation, if one considers the parametrizations used in the previous case for $q$ and $\om$, it is easy to show that
\be
\lb{e9}
\int_{0}^{\infty} \om^{k} \mathpzc{Y}_{r}(\om;q) \, d \om = q^{- k(k+r)} ,
\ee
which leads us to validate the integral representation
\be
\lb{e10}
\int_{0}^{\infty} \mathscr{G}_{n}(q^{n+r} z \om;q) \mathpzc{Y}_{r}(\om;q) \, d \om = \prz_{n}(z;q) .
\ee
In such a case, contrasting with the relation $\mathscr{G}_{n}(z;q) = \mathscr{H}_{n}(z;q^{-1})$, the transformation kernel 
$\mathpzc{Y}_{r}(\om;q)$ permits us to identify a new link between the SW and RS polynomials through the integral representation
(\ref{e10}). Note that equations (\ref{e6}) and (\ref{e10}) represent, in particular, two new additional results within a wide 
variety of formal properties obtained by Carlitz \cite{Carlitz} for $\{ \prz_{n}(z;q) \}_{n \in \mathbb{N}}$. Next, we will investigate 
a special family of $q$-orthogonalized functions (here named as the RS functions) with the aim of obtaining a set of mathematical 
properties which leads us to formally characterize its associated algebraic structure.

%%%%%%%%%%%%%%%%%%%%%%%%%%%%%%%%%%%%%%%%%%%%%%%%%%%%%%%%%%%%%%%%%%%%%%%%%%%%%%%%%%%%%%%%%%%%%%%%%%%%%%%%%%%%%%%%%%%%%%%%%%%%%%%%%%%%%%%%
\section{Algebraic properties of the RS functions}
%%%%%%%%%%%%%%%%%%%%%%%%%%%%%%%%%%%%%%%%%%%%%%%%%%%%%%%%%%%%%%%%%%%%%%%%%%%%%%%%%%%%%%%%%%%%%%%%%%%%%%%%%%%%%%%%%%%%%%%%%%%%%%%%%%%%%%%%

In this section we establish some preliminary mathematical results inherent to the RS functions. As a first step, we obtain a decomposition 
formula for the Szeg\"{o} measure in the complex plane which leads us to define a weight function $\erz(\varphi;q)$ related to our object 
of study. In the following step, we construct explicitly the realizations of raising and lowering operators for such functions by means of 
a specific definition of the Jackson's $q$-derivative, as well as discuss the implications of this algebraic approach.  

%%%%%%%%%%%%%%%%%%%%%%%%%%%%%%%%%%%%%%%%%%%%%%%%%%%%%%%%%%%%%%%%%%%%%%%%%%%%%%%%%%%%%%%%%%%%%%%%%%%%%%%%%%%%%%%%%%%%%%%%%%%%%%%%%%%%%%%%
\subsection{Preliminaries}
%%%%%%%%%%%%%%%%%%%%%%%%%%%%%%%%%%%%%%%%%%%%%%%%%%%%%%%%%%%%%%%%%%%%%%%%%%%%%%%%%%%%%%%%%%%%%%%%%%%%%%%%%%%%%%%%%%%%%%%%%%%%%%%%%%%%%%%%

The measure employed in equation (\ref{e4}) to orthogonalize the RS polynomials admits an expression of the form
\be
\lb{e11}
\vartheta_{3} \lpar \frac{\varphi}{2} \biggl| q^{\half} \rpar = \sum_{\ell \in \mathbb{Z}} q^{\half \ell^{2}} e^{\nc \ell \varphi} = 
1 + 2 \sum_{\ell \in \mathbb{N}^{\ast}} q^{\half \ell^{2}} \cos (\ell \varphi) .
\ee
Since equation (\ref{e11}) defines a strictly positive function on the interval $\varphi \in [-\pi,\pi]$ for any $q \in (0,1)$, let us 
properly employ the {\sl addition-formula} \cite{WW}
\bd
\fl \qquad \vartheta_{3}(x+y|\pq) \vartheta_{3}(x-y|\pq) \vartheta_{3}^{2}(0|\pq) = \vartheta_{3}^{2}(x|\pq) \vartheta_{3}^{2}(y|\pq) + 
\vartheta_{1}^{2}(x|\pq) \vartheta_{1}^{2}(y|\pq)
\ed
with $2 \pi$-period for $x=y=\case{\varphi}{4}$ and $\pq = q^{\half}$, in order to obtain
\bd
\vartheta_{3} \lpar \frac{\varphi}{2} \biggl| q^{\half} \rpar \vartheta_{3}^{3} \Bigl( 0 \Bigl| q^{\half} \Bigr) = \vartheta_{3}^{4} 
\lpar \frac{\varphi}{4} \biggl| q^{\half} \rpar + \vartheta_{1}^{4} \lpar \frac{\varphi}{4} \biggl| q^{\half} \rpar .
\ed
So, if one considers the complex function (up to a phase factor)
\be
\lb{e12}
\fl \qquad \erz(\varphi;q) \coloneq \lbk \vartheta_{3} \Bigl( 0 \Bigl| q^{\half} \Bigr) \rbk^{- \frac{3}{2}} \lbk \vartheta_{3}^{2} 
\lpar \frac{\varphi}{4} \biggl| q^{\half} \rpar + \nc \vartheta_{1}^{2} \lpar \frac{\varphi}{4} \biggl| q^{\half} \rpar \rbk ,
\ee
it is immediate to verify that (\ref{e11}) can also be written as a product of $\erz(\varphi;q)$ by its respective complex conjugate. 
Figure \ref{measure} shows the plots of the equipotential curves related to (a) $\mathrm{Re} \lbk \erz(\varphi;q) \rbk$ and (b) $\mathrm{Im} 
\lbk \erz(\varphi;q) \rbk$ for the variables $\varphi$ and $q$ within the intervals $[- \pi,\pi ]$ and $(0,1)$, respectively. Note 
that, in particular, plot (a) presents a peak at the point $\varphi = 0$ and $q \rightarrow 1$, this behaviour being the same of that observed 
in our numerical investigations for the equipotential curves associated with the Szeg\"{o} measure. In counterpart, plot (b) has two symmetric 
peaks located at the points $\varphi = - \pi,\pi$ and $q \rightarrow 0$, such distinct behaviour being now explained due to the presence of the 
Jacobi $\vartheta_{1}$-function in equation (\ref{e12}). Consequently, the product $\erz(\varphi;q) \erz^{\ast}(\varphi;q)$ not only retains 
the pattern verified in plot (a) but also confirms an important property for this integration measure, that is, its decomposition into the 
complex plane. Next, we define properly our object of study.
%%%%%%%%%%%%%%%%%%%%%%%%%%%%%%%%%%%%%%%%%%%%%%%%%%%%%%%%%%%%%%%%%%%%%%%%%%%%%%%%%%%%%%%%%%%%%%%%%%%%%%%%%%%%%%%%%%%%%%%%%%%%%%%%%%%%%%%%%
\begin{figure}[!t]
\centering
\begin{minipage}[b]{0.45\linewidth}
\includegraphics[width=\textwidth]{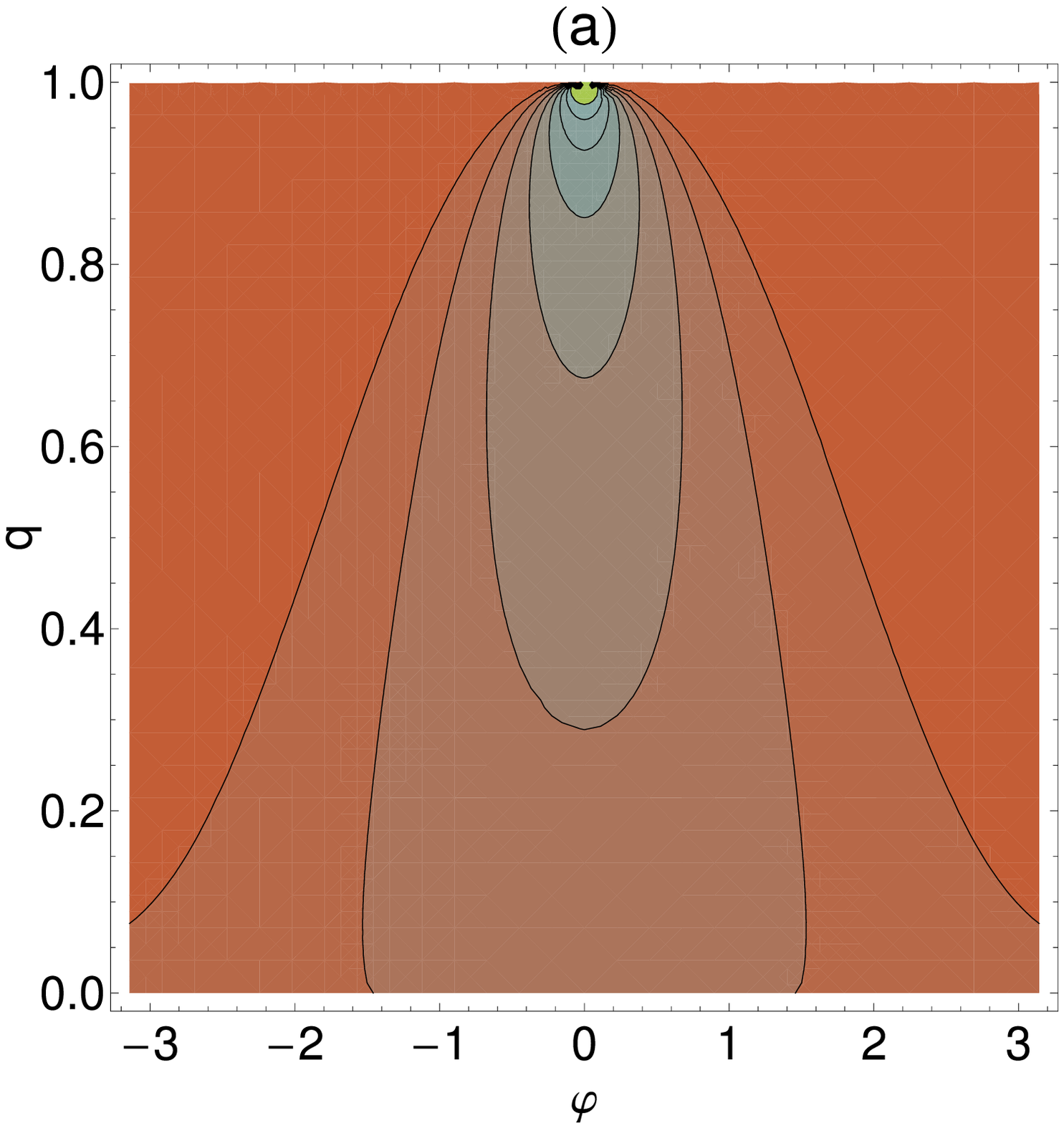}
\end{minipage} \hfill
\begin{minipage}[b]{0.45\linewidth}
\includegraphics[width=\textwidth]{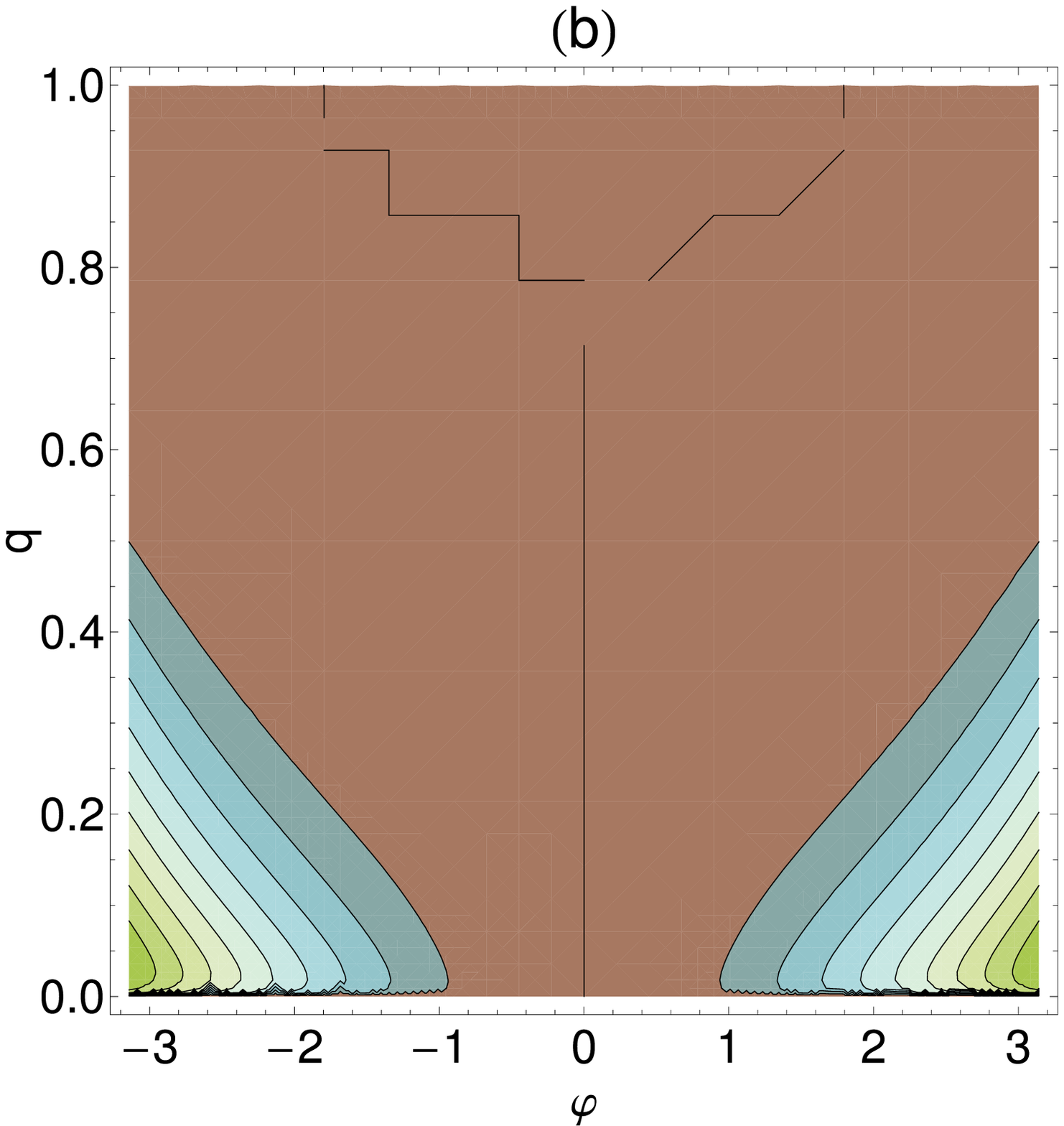}
\end{minipage}
\caption{\lb{measure} (Color online) Plots of equipotential curves associated with the (a) real and (b) imaginary parts of the complex 
function $\erz(\varphi;q)$ for $\varphi \in [-\pi,\pi]$ and $q \in (0,1)$. The different patterns observed in both pictures are directly 
related to the distinct behaviours of the Jacobi $\vartheta_{3}$- and $\vartheta_{1}$-functions used to decompose the Szeg\"{o} measure 
in the product $\erz(\varphi;q) \erz^{\ast}(\varphi;q)$.} 
\end{figure}
%%%%%%%%%%%%%%%%%%%%%%%%%%%%%%%%%%%%%%%%%%%%%%%%%%%%%%%%%%%%%%%%%%%%%%%%%%%%%%%%%%%%%%%%%%%%%%%%%%%%%%%%%%%%%%%%%%%%%%%%%%%%%%%%%%%%%%%%%%

%%%%%%%%%%%%%%%%%%%%%%%%%%%%%%%%%%%%%%%%%%%%%%%%%%%%%%%%%%%%%%%%%%%%%%%%%%%%%%%%%%%%%%%%%%%%%%%%%%%%%%%%%%%%%%%%%%%%%%%%%%%%%%%%%%%%%%%%
\subsubsection*{{\bf Definition.}}
%%%%%%%%%%%%%%%%%%%%%%%%%%%%%%%%%%%%%%%%%%%%%%%%%%%%%%%%%%%%%%%%%%%%%%%%%%%%%%%%%%%%%%%%%%%%%%%%%%%%%%%%%%%%%%%%%%%%%%%%%%%%%%%%%%%%%%%%

{\it Let us initially introduce the RS functions by means of the product
\be
\lb{e13}
\fl \qquad \quad \Psi_{n}^{\inrs}(z;q) \coloneq \lbk \frac{q^{n}}{2 \pi (q;q)_{n}} \rbk^{\half} \prz_{n}(z;q) \mathscr{M}(z;q) = 
\mathscr{R}_{n}(z;q) \mathscr{M}(z;q) ,
\ee
where $\mathscr{R}_{n}(z;q)$ denotes the RS polynomials (at least of the normalization coefficient), and 
\bd
\mathscr{M}(z;q) = \lbk \mathfrak{F} \Bigl( - q^{- \half};q \Bigr) \rbk^{- \case{3}{2}} \lbk \mathfrak{F}^{2}(z;q) + \nc 
\mathfrak{G}^{2}(z;q) \rbk  
\ed
represents a weight function with
\brr
\lb{e14}
\mathfrak{F}(z;q) &=& \sum_{\ell \in \mathbb{Z}} (- \nc)^{\ell} q^{\half \ell \lpar \ell + \half \rpar} z^{\case{\ell}{2}} , \\
\lb{e15}
\mathfrak{G}(z;q) &=& - \sum_{\ell \in \mathbb{Z}} \nc^{\ell + \half} q^{\half \lpar \ell + \half \rpar (\ell + 1)} z^{\half \lpar 
\ell + \half \rpar} .
\err
Thus, for $z = - q^{- \half} e^{\nc \varphi}$ fixed, it is immediate to see that $\mathscr{M} \Bigl(- q^{- \half} e^{\nc \varphi};q \Bigr) 
\equiv \mathscr{E}(\varphi;q)$ since equations (\ref{e14}) and (\ref{e15}) coincide, respectively, with the $\vartheta_{3}$- and
$\vartheta_{1}$-functions. In addition, this particular parametrization also permits us to verify that
\be
\lb{e16}
\int_{-\pi}^{\pi} {\Psi_{m}^{\inrs}}^{\ast} \Bigl(- q^{- \half} e^{\nc \varphi};q \Bigr) \Psi_{n}^{\inrs} \Bigl(- q^{- \half} 
e^{\nc \varphi};q \Bigr) d \varphi = \delta_{mn} .
\ee
Hence, $\{ \Psi_{n}^{\inrs}(z;q) \}_{n \in \mathbb{N}}$ features a set of well-defined functions into the complex plane which are
orthogonalized on the unit circle.}

The proof of the completeness relation associated with $\{ \Psi_{n}^{\inrs}(z;q) \}_{n \in \mathbb{N}}$ follows basically the formal 
mathematical treatment sketched in Refs. \cite{Ismail,Atakishiyev}. For this task, we first define the bilinear kernel
\be
\lb{e17}
\mathrm{K}_{\eps}(w,z;q) \coloneq \sum_{n \in \mathbb{N}} \eps^{n} \, {\Psi_{n}^{\inrs}}^{\ast}(w;q) \Psi_{n}^{\inrs}(z;q) \qquad 
| \eps | < 1
\ee
which can be evaluated by means of the auxiliary relation \cite{Andrews}
\bd
\sum_{n \in \mathbb{N}} \prz_{n}(w^{\ast};q) \prz_{n}(z;q) \, \frac{(q \eps)^{n}}{(q;q)_{n}} = \frac{\lpar q^{2} \eps^{2} w^{\ast} z ;
q \rpar_{\infty}}{\lpar q \eps w^{\ast} z, q \eps w^{\ast}, q \eps z, q \eps; q \rpar_{\infty}} , 
\ed
namely\footnote{The notation $\lpar a_{1}, a_{2}, \ldots, a_{r}; q \rpar_{\infty} \equiv (a_{1};q)_{\infty} \cdot (a_{2};q)_{\infty} 
\cdot \ldots \cdot (a_{r};q)_{\infty}$ here employed represents the generalized $q$-Pochhammer symbol with $\{ a_{1}, a_{2}, \ldots, 
a_{r} \} \in \mathbb{C}$.}
\bd
\mathrm{K}_{\eps}(w,z;q) = \frac{\lpar q^{2} \eps^{2} w^{\ast} z; q \rpar_{\infty} \mathscr{M}^{\ast}(w;q) \mathscr{M}(z;q)}
{2 \pi \lpar q \eps w^{\ast} z, q \eps w^{\ast}, q \eps z, q \eps; q \rpar_{\infty}} .
\ed
Note that for $z = - q^{- \half} e^{\nc \varphi}$ and $w = - q^{- \half} e^{\nc \beta}$, $\mathrm{K}_{\eps}(w,z;q)$ is expressed as
\be
\lb{e18}
\fl \qquad \mathrm{K}_{\eps} (\beta, \varphi; q) = \frac{\Bigl( q \eps^{2} e^{\nc (\varphi - \beta)};q \Bigr)_{\infty} \, \mathscr{E}^{\ast}
(\beta;q) \mathscr{E}(\varphi;q)}{2 \pi \Bigl( \eps e^{\nc (\varphi - \beta)}, - q^{\half} \eps e^{- \nc \beta}, - q^{\half} \eps 
e^{\nc \varphi}, q \eps; q \Bigr)_{\infty}} .
\ee
Besides, if one takes into account the orthogonality relation (\ref{e16}), it is immediate to show that equation (\ref{e18}) obeys the
following properties:
\brr
\lb{e19}
\fl \qquad \quad \int_{-\pi}^{\pi} \mathrm{K}_{\eps}(\beta, \varphi; q) \Psi_{n}^{\inrs} \Bigl(- q^{- \half} e^{\nc \beta};q \Bigr) 
d \beta = \eps^{n} \Psi_{n}^{\inrs} \Bigl(- q^{- \half} e^{\nc \varphi};q \Bigr) , \\
\lb{e20}
\fl \qquad \quad \int_{-\pi}^{\pi} \mathrm{K}_{\eps}(\beta, \varphi; q) \mathrm{K}_{\eps^{\prime}}(\gamma, \beta; q) d \beta = 
\mathrm{K}_{\eps \eps^{\prime}}(\gamma, \varphi; q) .
\err
Thus, any well-behaved (or at least piecewise continuous) function $F(z)$ can now be properly expanded in terms of the complete set 
of functions $\{ \Psi_{n}^{\inrs}(z;q) \}_{n \in \mathbb{N}}$.

Next, we discuss certain relevant additional points associated with the definition proposed for $\Psi_{n}^{\inrs}(z;q)$, which will be useful 
in the descriptive process of its formal properties.
\begin{description}
\item[(i)] Guided by the analogy with the usual harmonic oscillator (HO) on the line, where the normalized wavefunction 
\cite[page 151]{Ballentine}
\bd
\Psi_{n}^{\inho}(x;\alf) = \lpar \frac{\alf}{\pi^{1/2} 2^{n} n!} \rpar^{\half} H_{n}( \alf x ) \, e^{- \half (\alf x)^{2}} \qquad
(n \in \mathbb{N})
\ed
is written in terms of the Hermite polynomials $H_{n}(\alf x)$ and the Gaussian weight function $\exp \lbk - \half (\alf x)^{2} \rbk$ 
(note that $\alf$ is the HO width and also acts as a controlling parameter in this case), we may guess that (\ref{e13}) plays a similar 
role and the angular density function $\bigl| \Psi_{n}^{\inrs} \bigl( - q^{- \half} e^{\nc \varphi};q \bigr) \bigr|^{2}$ is, as a matter 
of fact, a good candidate in describing a phase distribution for a $q$-deformed HO, with $q$ being a parameter that controls the 
distribution width, and therefore responsible for {\em squeezing effects} \cite{Galetti}. In order to reinforce such an argument, figure 
\ref{RS-functions} shows this particular phase distribution as a function of the angular variable $\varphi \in [- \pi,\pi ]$ and different
values of $q$, where the excitation degree $n$ of the $q$-deformed HO is restricted into the closed interval $[0,4]$ --- see figures 2(a)-2(e).
%%%%%%%%%%%%%%%%%%%%%%%%%%%%%%%%%%%%%%%%%%%%%%%%%%%%%%%%%%%%%%%%%%%%%%%%%%%%%%%%%%%%%%%%%%%%%%%%%%%%%%%%%%%%%%%%%%%%%%%%%%%%%%%%%%%%%%%%%
\begin{figure}[!t]
\centering
\begin{minipage}[b]{0.4\linewidth}
\includegraphics[width=\textwidth]{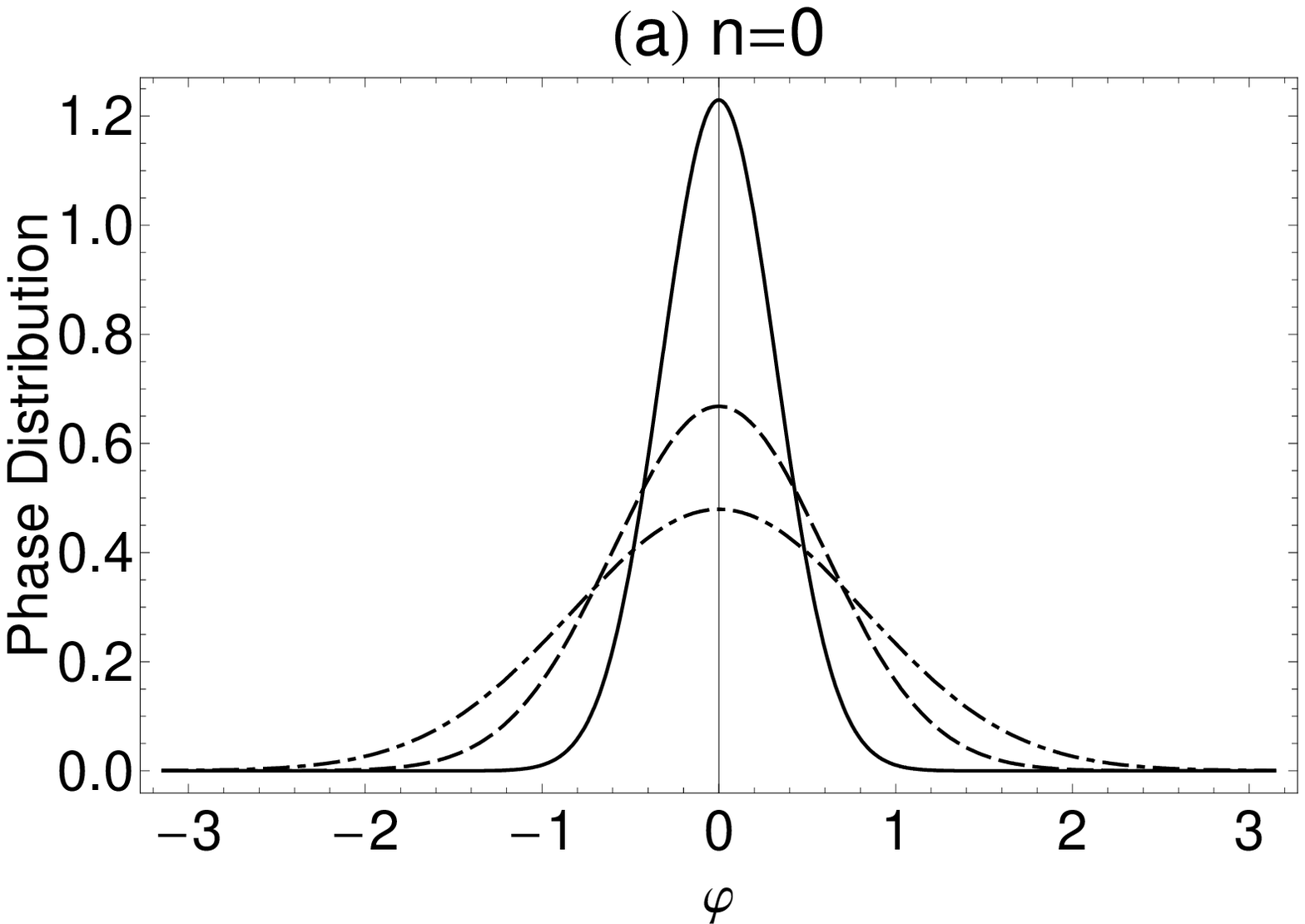}
\end{minipage} \hfill
\begin{minipage}[b]{0.4\linewidth}
\includegraphics[width=\textwidth]{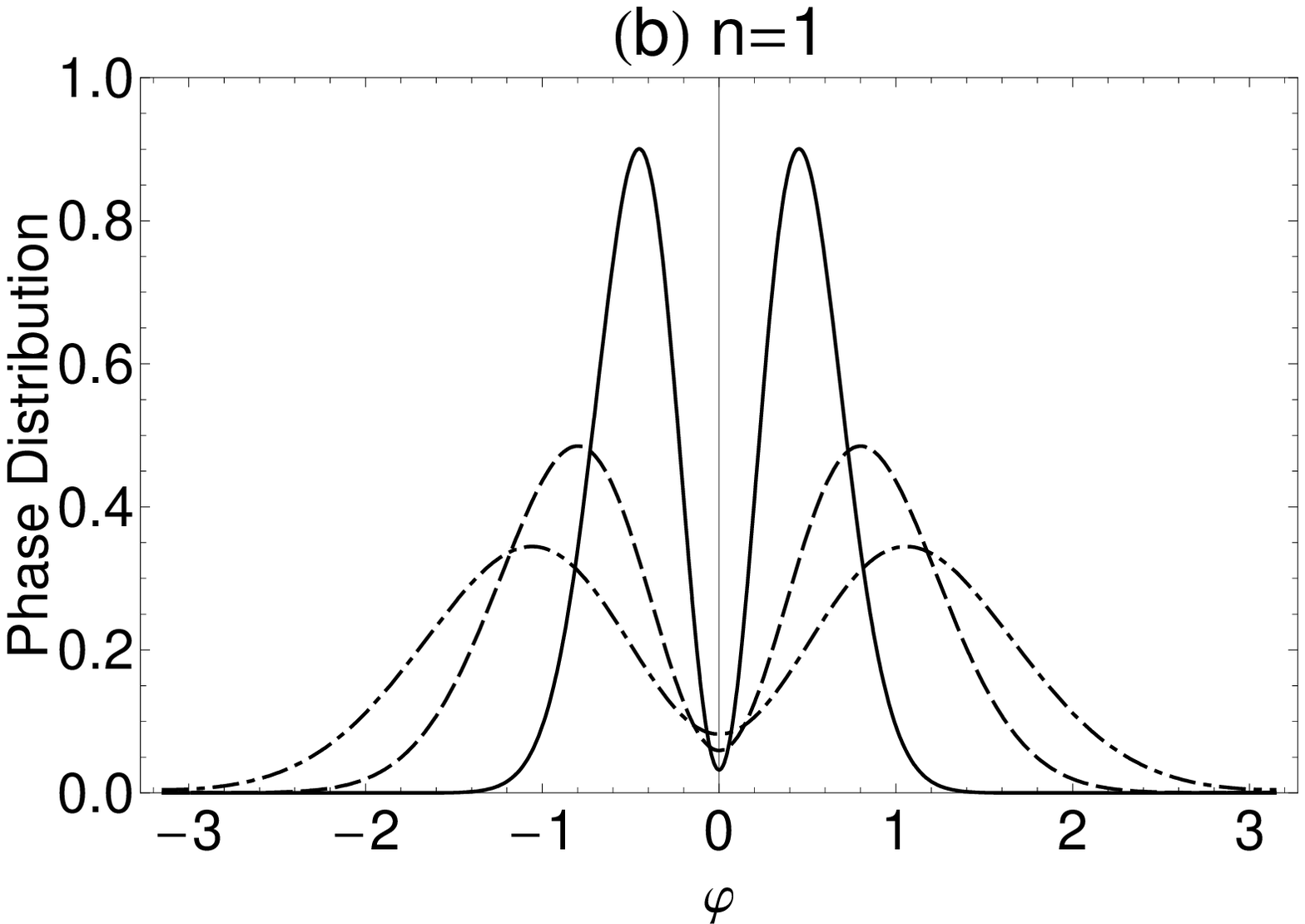}
\end{minipage} \hfill
\begin{minipage}[b]{0.4\linewidth}
\includegraphics[width=\textwidth]{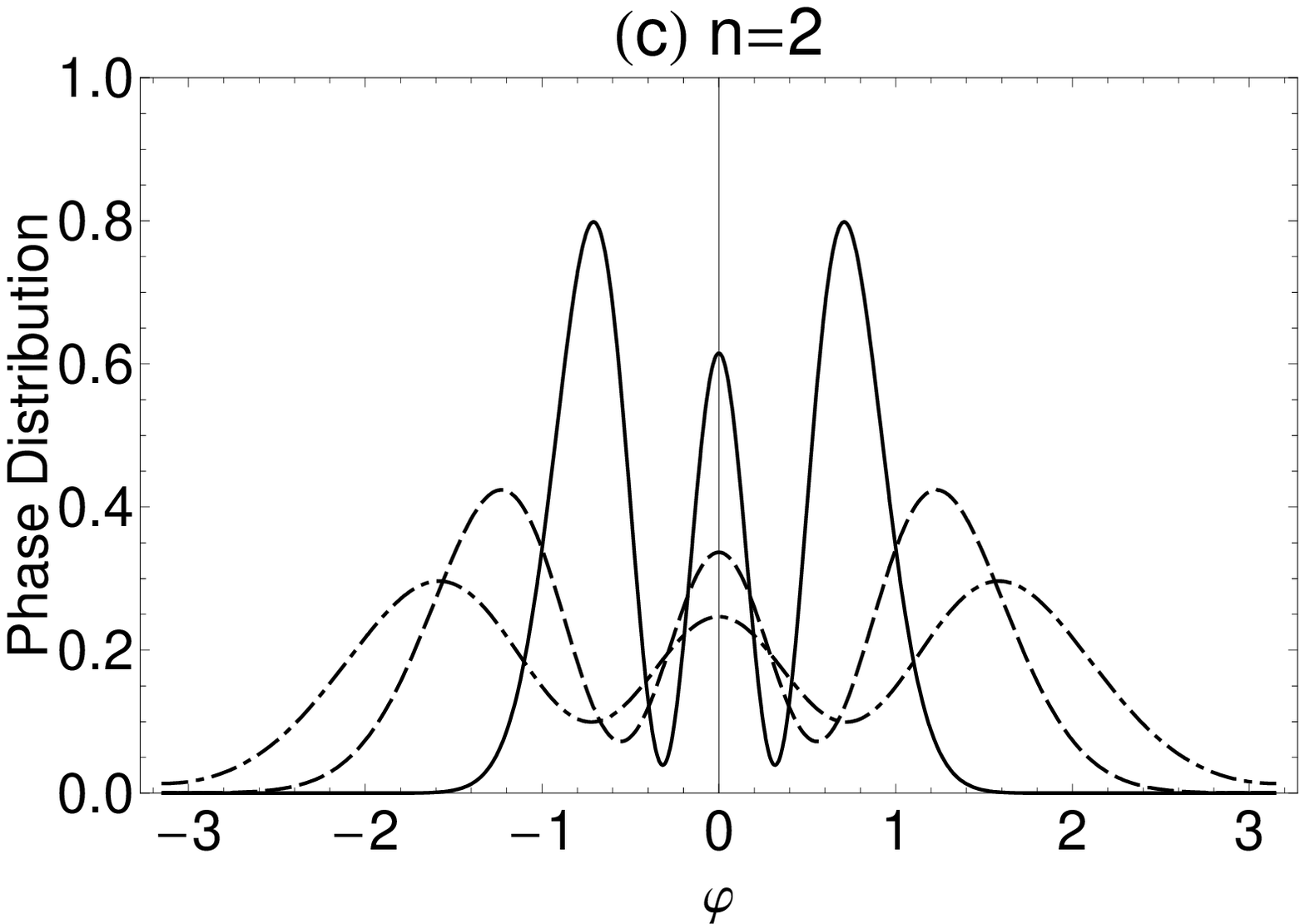}
\end{minipage} \hfill
\begin{minipage}[b]{0.4\linewidth}
\includegraphics[width=\textwidth]{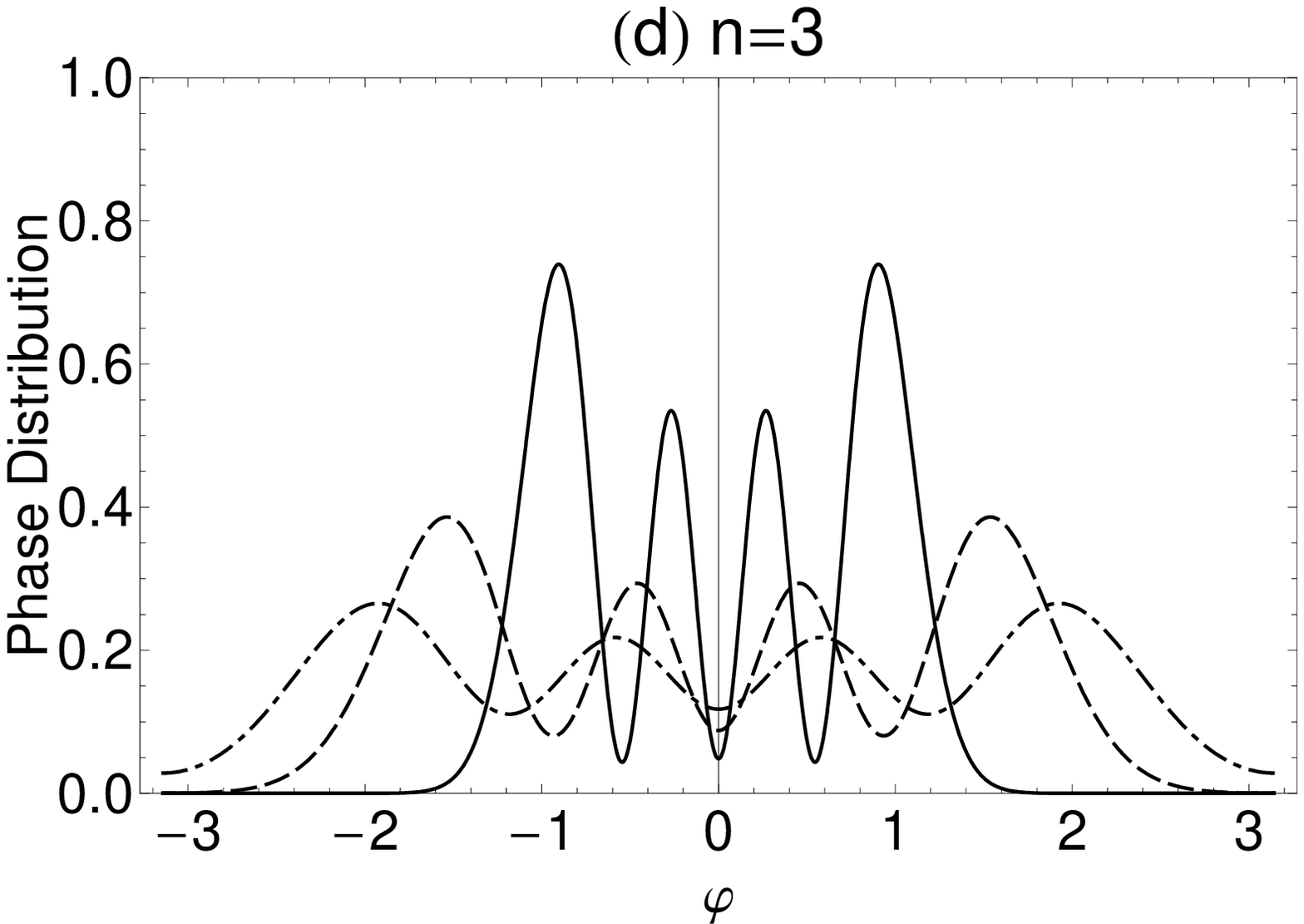}
\end{minipage} \hfill
\begin{minipage}[b]{0.4\linewidth}
\includegraphics[width=\textwidth]{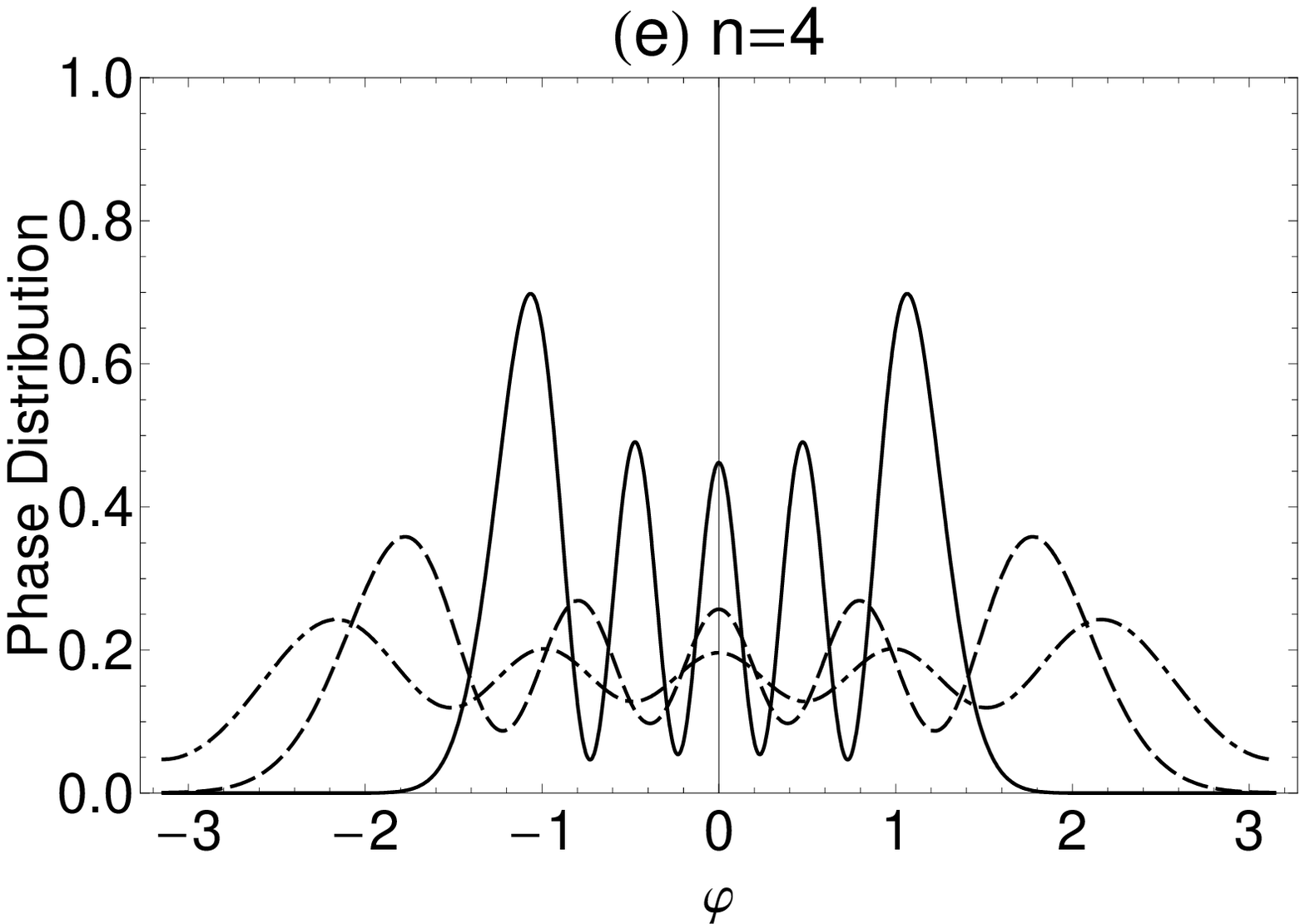}
\end{minipage}
\caption{\lb{RS-functions} Plots of $\bigl| \Psi_{n}^{\inrs} \bigl( - q^{- \half} e^{\nc \varphi};q \bigr) \bigr|^{2}$ as a function of the
angle $\varphi \in [- \pi,\pi ]$ with $n \in [0,4]$ and different values of $q$, such as, for example, $q = 0.5$ (dot-dashed line), $0.7$ 
(dashed line) and $0.9$ (solid line). In particular, these plots show how the curves associated with each $n$ of a $q$-deformed HO are 
affected by the parameter $q \in (0,1)$. It is worth noticing that such phase distribution is a well-behaved function for both variables $n$ 
and $\varphi$ defined on a compact support, and it has a specular reflection (or symmetric behaviour) at the origin $\varphi = 0$.} 
\end{figure}
%%%%%%%%%%%%%%%%%%%%%%%%%%%%%%%%%%%%%%%%%%%%%%%%%%%%%%%%%%%%%%%%%%%%%%%%%%%%%%%%%%%%%%%%%%%%%%%%%%%%%%%%%%%%%%%%%%%%%%%%%%%%%%%%%%%%%%%%%%
%
\item[(ii)] The three-term recurrence relation (\ref{e3}) can be promptly adapted for $\{ \mathscr{R}_{n}(z;q) \}_{n \in \mathbb{N}}$ as 
follows:
\bd
\fl \quad \mathscr{R}_{n+1}(z;q) = \lpar \frac{q}{1-q^{n+1}} \rpar^{\half} \lbr (1 + z) \mathscr{R}_{n}(z;q) - \lbk q (1-q^{n}) 
\rbk^{\half} z \, \mathscr{R}_{n-1}(z;q) \rbr .
\ed
Furthermore, after some calculations based on the results obtained in Ref. \cite{Carlitz} for the RS polynomials, it is easy to reach
the additional relations 
\brr
\fl & \mathscr{R}_{n}(z;q) - \mathscr{R}_{n}(qz;q) = \lbk q \lpar 1 - q^{n} \rpar \rbk^{\half} z \, \mathscr{R}_{n-1}(z;q) , \nn \\
\fl & \mathscr{R}_{n}(qz;q) - q^{n} \mathscr{R}_{n}(z;q) = \lbk q \lpar 1 - q^{n} \rpar \rbk^{\half} \mathscr{R}_{n-1}(qz;q) , \nn \\
\fl & \mathscr{R}_{n}(qz;q) - \mathscr{R}_{n}(q^{2} z;q) = \lbk q \lpar 1 - q^{n} \rpar \rbk^{\half} qz \, \mathscr{R}_{n-1}(qz;q) , \nn \\
\fl & \mathscr{R}_{n}(z;q) - \mathscr{R}_{n}(q^{2} z;q) = q \lpar 1 - q^{n} \rpar z \, \mathscr{R}_{n}(z;q) + \lbk q (1-q^{n}) 
\rbk^{\half} (1 - qz) z \, \mathscr{R}_{n-1}(z;q) . \nn 
\err
In particular, these identities show how certain recurrence relations are modified by the scaling factors $q$ and $q^{2}$.
\item[(iii)] This behaviour can also be verified for the weight function $\mathscr{M}(z;q)$, namely, scaling factors involving odd 
and even powers of the parameter $q$ also modify equations (\ref{e14}) and (\ref{e15}) --- or, in other words, they change the 
quasi-period of the Jacobi theta functions. In such cases, if one considers $r$ an integer number, we obtain
\brr
\fl \quad \mbox{({\bf a}) $q^{2r}$-case} \nn \\
\fl \qquad \quad \mathfrak{F}(q^{2r} z;q) &=& \nc^{r} q^{- \half r \lpar r + \half \rpar} z^{- \case{r}{2}} \mathfrak{F}(z;q) \nn \\
\fl \qquad \quad \mathfrak{G}(q^{2r} z;q) &=& (- \nc)^{r} q^{- \half r \lpar r + \half \rpar} z^{- \case{r}{2}} \mathfrak{G}(z;q) \nn \\
\fl \qquad \quad \mathscr{M}(q^{2r} z;q) &=& (-1)^{r} q^{- r \lpar r + \half \rpar} z^{-r} \mathscr{M}(z;q) \nn
\err
and
\brr
\fl \quad \mbox{({\bf b}) $q^{2r+1}$-case} \nn \\
\fl \qquad \quad \mathfrak{F}(q^{2r+1} z;q) &=& \nc^{r + \half} q^{- \half \lpar r + \half \rpar (r+1)} z^{- \half \lpar r + \half \rpar}
\mathfrak{F}(z;q) \nn \\
\fl \qquad \quad \mathfrak{G}(q^{2r+1} z;q) &=& (- \nc)^{r + \half} q^{- \half \lpar r + \half \rpar (r+1)} z^{- \half \lpar r + \half \rpar}
\mathfrak{G}(z;q) \nn \\
\fl \qquad \quad \mathscr{M}(q^{2r+1} z;q) &=& (-1)^{r + \half} q^{- \lpar r + \half \rpar (r+1)} z^{- \lpar r + \half \rpar} 
\mathscr{M}^{\ast}(z;q) . \nn
\err
Note that scaling factors containing odd powers of $q$ change the phase of the complex function $\mathscr{M}(z;q)$. Consequently, this
result will bring some implications for the algebraic properties of $\Psi_{n}^{\inrs}(z;q)$ and also will be responsible for the small
change made in the usual definition of the Jackson's $q$-derivative \cite{Gasper,Vilenkin}.
\item[(iv)] Adiga and coworkers \cite[page 29]{Ramanujan} have established and proved several properties originated from the Ramanujan's 
theorems on
\bd
f(a,b) \coloneq \sum_{\ell \in \mathbb{Z}} \, a^{\half \ell (\ell + 1)} b^{\half \ell (\ell - 1)} ,
\ed
where $| ab | < 1$ (we retain the original notation). Since $a$ and $b$ denote two complex variables in such a case, if one sets 
$a = q^{\half} e^{\nc \varphi}$ and $b = q^{\half} e^{- \nc \varphi}$, it is immediate to verify that $f(a,b)$ coincides exactly 
with equation (\ref{e11}), \ie,
\bd
\fl \quad f \Bigl( q^{\half} e^{\nc \varphi}, q^{\half} e^{- \nc \varphi} \Bigr) = \Bigl| \mathscr{M} \Bigl(- q^{- \half} 
e^{\nc \varphi};q \Bigr) \Bigr|^{2} = | \mathscr{E}(\varphi;q) |^{2} = \vartheta_{3} \lpar \frac{\varphi}{2} \biggl| q^{\half} 
\rpar .
\ed
This particular connection allows us to increase the number of properties related to the Szeg\"{o} measure (if one compares with those
obtained in this work), as well as to derive new scaling relations.
\end{description}

%%%%%%%%%%%%%%%%%%%%%%%%%%%%%%%%%%%%%%%%%%%%%%%%%%%%%%%%%%%%%%%%%%%%%%%%%%%%%%%%%%%%%%%%%%%%%%%%%%%%%%%%%%%%%%%%%%%%%%%%%%%%%%%%%%%%%%%%%
\subsection{Jackson's $q$-derivative}
%%%%%%%%%%%%%%%%%%%%%%%%%%%%%%%%%%%%%%%%%%%%%%%%%%%%%%%%%%%%%%%%%%%%%%%%%%%%%%%%%%%%%%%%%%%%%%%%%%%%%%%%%%%%%%%%%%%%%%%%%%%%%%%%%%%%%%%%%

The next step consists in introducing a particular choice for the Jackson's $q$-derivative through the action of a certain $q$-differential 
operator $\mathpzc{D}_{q^{2}}$ on an arbitrary complex function $\phi(z;q)$ as follows:\footnote{It is important to stress that exist
different versions of the Jackson's $q$-derivative in the literature covering a wide range of applications in specific scenarios of 
mathematics and physics. For instance, Gelfand and coworkers \cite{Gelfand} have proposed a two-parameter $q$-differential operator 
$\mathpzc{D}_{r,s}$ whose action on $\phi(z;q)$ obeys the mathematical prescription
\bd
\mathpzc{D}_{r,s} \phi(z;q) \coloneq \frac{\phi(rz;q) - \phi(sz;q)}{z(r-s)} .
\ed
Note that $\mathpzc{D}_{q^{2}}$ represents a particular case of $\mathpzc{D}_{r,s}$ since $\mathpzc{D}_{q^{2}} \equiv \mathpzc{D}_{1,q^{2}}$.}
\be
\lb{e21}
\mathpzc{D}_{q^{2}} \phi(z;q) \coloneq \frac{\phi(z;q) - \phi(q^{2} z;q)}{z (1 - q^{2})} .
\ee
It is worth mentioning that some useful rules of $q$-differentiation, analogous to those verified for ordinary differentiation, can also be 
directly obtained in the context of $q$-calculus \cite{Jackson1,Jackson2}. Among them, let us focus on two basic rules which have an important 
role in our calculations, that is, the sum rule
\be
\lb{e22}
\mathpzc{D}_{q^{2}} \lbk \phi_{1}(z;q) + \phi_{2}(z;q) \rbk = \mathpzc{D}_{q^{2}} \phi_{1}(z;q) + \mathpzc{D}_{q^{2}} \phi_{2}(z;q)
\ee
and the $q$-version of the Leibnitz rule
\brr
\lb{e23}
\fl \qquad \mathpzc{D}_{q^{2}} \lbk \phi_{1}(z;q) \phi_{2}(z;q) \rbk &=& \lbk \mathpzc{D}_{q^{2}} \phi_{1}(z;q) \rbk \phi_{2}(z;q) + 
\phi_{1}(q^{2} z;q) \lbk \mathpzc{D}_{q^{2}} \phi_{2}(z;q) \rbk \nn \\
&=& \lbk \mathpzc{D}_{q^{2}} \phi_{1}(z;q) \rbk \phi_{2}(q^{2} z;q) + \phi_{1}(z;q) \lbk \mathpzc{D}_{q^{2}} \phi_{2}(z;q) \rbk .
\err

As a first task, let us determine the action of the $q$-differential operator $\mathpzc{D}_{q^{2}}$ on the polynomial $\mathscr{R}_{n}(z;q)$,
that is, 
\be
\lb{e24}
\fl \quad \mathpzc{D}_{q^{2}} \mathscr{R}_{n}(z;q) = \frac{q}{1+q} [ n ]_{q} \, \mathscr{R}_{n}(z;q) + \lpar \frac{q}{1-q} \rpar^{\half}
\frac{1-qz}{1+q} [ n ]_{q}^{1/2} \mathscr{R}_{n-1}(z;q)
\ee
where $[ n ]_{q} \coloneq \case{1-q^{n}}{1-q}$ stands for the $q$-number \cite{Jackson1,Jackson2}. The second task then consists in 
finding out an expression for $\mathpzc{D}_{q^{2}} \mathscr{M}(z;q)$ through the formal results obtained in the previous discussion 
about scaling factors,
\be
\lb{e25}
\mathpzc{D}_{q^{2}} \mathscr{M}(z;q) = \frac{1 + q^{\case{3}{2}} z}{q^{\case{3}{2}} z^{2} (1-q^{2})} \mathscr{M}(z;q) .
\ee
Consequently, the action of $\mathpzc{D}_{q^{2}}$ on $\Psi_{n}^{\inrs}(z;q)$ can now be promptly obtained with the help of the
$q$-Leibnitz rule, \ie,
\bd
\fl \qquad \mathpzc{D}_{q^{2}} \Psi_{n}^{\inrs}(z;q) = \lbk \mathpzc{D}_{q^{2}} \mathscr{R}_{n}(z;q) \rbk \mathscr{M}(q^{2} z;q) +
\mathscr{R}_{n}(z;q) \lbk \mathpzc{D}_{q^{2}} \mathscr{M}(z;q) \rbk .
\ed
In this way, substituting equations (\ref{e24}) and (\ref{e25}) into this expression and after some minor adjustments in our calculations,
we finally get the relations
\bd
\fl \mathpzc{D}_{q^{2}} \Psi_{n}^{\inrs}(z;q) = \frac{1 - qz \Bigl( 1 - q^{\half} - q^{n} \Bigr)}{q^{\case{3}{2}} z^{2} (1-q^{2})}
\Psi_{n}^{\inrs}(z;q) - \lpar \frac{[n]_{q}}{1-q} \rpar^{\half} \frac{1-qz}{qz(1+q)} \Psi_{n-1}^{\inrs}(z;q)
\ed
and
\bd
\fl \mathpzc{D}_{q^{2}} \Psi_{n}^{\inrs}(z;q) = - \frac{1 - q \Bigl( z + q^{\half} + q^{n} \Bigr)}{q^{\case{3}{2}} z (1-q^{2})}
\Psi_{n}^{\inrs}(z;q) + \lpar \frac{[n+1]_{q}}{1-q} \rpar^{\half} \frac{1-qz}{q^{2} z^{2} (1+q)} \Psi_{n+1}^{\inrs}(z;q)
\ed
for any $q \in (0,1)$ and $n \in \mathbb{N}$. Note that both identities not only depend on the degrees $n$ and $n \mp 1$, but also 
preserve the phase of the orthogonalized RS functions --- this fact being, in particular, a direct consequence of the scaling relations 
derived for $\mathscr{M}(z;q)$. The comparison of these results leads us to verify that $\Psi_{n}^{\inrs}(z;q)$ satisfies a three-term
recurrence relation exactly equal to that obtained for $\mathscr{R}_{n}(z;q)$.

%%%%%%%%%%%%%%%%%%%%%%%%%%%%%%%%%%%%%%%%%%%%%%%%%%%%%%%%%%%%%%%%%%%%%%%%%%%%%%%%%%%%%%%%%%%%%%%%%%%%%%%%%%%%%%%%%%%%%%%%%%%%%%%%%%%%%%%%%
\subsection{Lowering and raising operators}
%%%%%%%%%%%%%%%%%%%%%%%%%%%%%%%%%%%%%%%%%%%%%%%%%%%%%%%%%%%%%%%%%%%%%%%%%%%%%%%%%%%%%%%%%%%%%%%%%%%%%%%%%%%%%%%%%%%%%%%%%%%%%%%%%%%%%%%%%

As a last step in our calculations, let us now construct the $q$-differential representations of the lowering and raising operators 
associated with $\{ \Psi_{n}^{\inrs}(z;q) \}_{n \in \mathbb{N}}$, as well as the respective representation for the $q$-deformed number 
operator. For this intent, both the results obtained for $\mathpzc{D}_{q^{2}} \Psi_{n}^{\inrs}(z;q)$ are taking into account in this process, 
asserting that
\be
\lb{e26}
\widehat{L}_{n}(z;q) \coloneq \frac{1 - qz \Bigl( 1 - q^{\half} - q^{n} \Bigr) - q^{\case{3}{2}} z^{2} (1 - q^{2}) \mathpzc{D}_{q^{2}}}
{[ q(1-q) ]^{\half} (1 - qz) z}
\ee
and
\be
\lb{e27}
\widehat{R}_{n}(z;q) \coloneq \frac{q^{\half} z \lbk 1 - q \Bigl( z + q^{\half} + q^{n} \Bigr) + q^{\case{3}{2}} z (1 - q^{2}) 
\mathpzc{D}_{q^{2}} \rbk}{(1-q)^{\half} (1 - qz)}
\ee
represent --- within the range of possibilities related to the definition of Jackson's $q$-derivative applied to the problem under scrutiny ---
two legitimate $q$-differential representations of the lowering $(\bop)$ and raising $(\bop^{\dagger})$ operators, respectively, whose actions 
on the RS functions result in\footnote{Although the $q$-differential representations $\widehat{L}_{n}(z;q)$ and $\widehat{R}_{n}(z;q)$ present 
an explicit dependence on the discrete variable $n$ (which certainly represents a possible drawback in our algebraic approach), it is worth 
stressing that the matrix elements of the raising and lowering operators can always be evaluated via RS-functions representation, and they are
sufficient to properly and uniquely determine such operators.}
\brr 
\bop \Psi_{n}^{\inrs}(z;q) & \equiv & \widehat{L}_{n}(z;q) \Psi_{n}^{\inrs}(z;q) = [n]_{q}^{1/2} \Psi_{n-1}^{\inrs}(z;q) , \nn \\
\bop^{\dagger} \Psi_{n}^{\inrs}(z;q) & \equiv & \widehat{R}_{n}(z;q) \Psi_{n}^{\inrs}(z;q) = [n+1]_{q}^{1/2} \Psi_{n+1}^{\inrs}(z;q) . \nn
\err
Moreover, if one defines
\be
\lb{e28}
\widehat{N}_{n}(z;q) \coloneq \widehat{R}_{n}(z;q) \widehat{L}_{n}(z;q)
\ee
as the number operator $\nop_{\indq}$ in its $q$-differential form, it is immediate to show that 
\bd
\nop_{\indq} \Psi_{n}^{\inrs}(z;q) \equiv \widehat{N}_{n}(z;q) \Psi_{n}^{\inrs}(z;q) = [n]_{q} \Psi_{n}^{\inrs}(z;q) .
\ed
Such $q$-differential representations can be considered as a particular realization (within a wide class of representations with 
different applications in mathematics and physics) of the IACK algebra \cite{Burban} here characterized by the commutation relations
\begin{itemize}
\item $q$-commutation relation $\lpar [ {\bf X},{\bf Y} ]_{q} \equiv {\bf X} {\bf Y} - q {\bf Y} {\bf X} \rpar$
\be
\lb{e29} 
\fl \quad [ \bop,\bop^{\dagger} ]_{q} = {\bf 1}, \quad [ \bop,\nop_{\indq} ]_{q} = \bop, \quad [ \nop_{\indq},\bop^{\dagger} ]_{q} = 
\bop^{\dagger}, 
\ee
\item standard commutation relation $\lpar [ {\bf X},{\bf Y} ] \equiv {\bf X} {\bf Y} - {\bf Y} {\bf X} \rpar$
\brr
\lb{e30}
\fl & \quad [ \bop,\bop^{\dagger} ] = {\bf 1}-(1-q) \nop_{\indq} , \quad [ \nop_{\indq},\bop ] = - [{\bf 1}-(1-q) \nop_{\indq}] 
\bop , \nn \\
\fl & \qquad \qquad \qquad [ \nop_{\indq},\bop^{\dagger} ] = \bop^{\dagger} [{\bf 1}-(1-q) \nop_{\indq}] . 
\err
\end{itemize}
Next, we will discuss certain relevant points related to the results previously obtained in this paragraph. 

Our first comment refers to the $q$-differential forms determined for the lowering, raising and number operators and their dependences
on the degree $n$, this fact being interpreted as a direct consequence of the definition employed in this work for the RS functions and 
its inherent properties -- indeed, the expressions for $\mathpzc{D}_{q^{2}} \Psi_{n}^{\inrs}(z;q)$ already bring such dependence. 
Besides, the energy eigenvalues \cite{Burban,Marchiolli}
\bd
E_{n} = \frac{2 - (1+q) q^{n}}{1 - q} \, E_{0} \qquad \lpar E_{0} = \hbar \om_{0}/2 \rpar
\ed
of this particular $q$-deformed HO present a nonlinear dependence on $n$ and its energy spectrum is not equally spaced, namely
\bd
\Delta_{n} \coloneq \frac{E_{n+1} - E_{n}}{E_{0}} = (1+q) q^{n} .
\ed
In principle, such arguments should be sufficient to explain qualitatively the functional forms of equations (\ref{e26}) and (\ref{e27}).

The second comment is related to the commutation relations (\ref{e29}) and (\ref{e30}), where it is clear that both relations are equivalent 
only in the contraction limit $q \rightarrow 1^{-}$ (in this limit, we recover the Heisenberg-Weyl algebra $\mathfrak{h}_{4}$); furthermore, 
analogous commutation relations were also obtained in Ref. \cite{Galetti} for the RS polynomials. As a last remark, let us mention that 
Floreanini and Vinet \cite{Floreanini} have developed a quantum-algebraic framework for a finite set of $q$-special functions, where the 
realizations of the underlying algebras are given in terms of operators acting on vector spaces of complex functions. In this sense, the 
results here discussed may represent an important contribution to that framework since it allows us to open a new promising chapter on the 
angular representations in quantum mechanics. 

%%%%%%%%%%%%%%%%%%%%%%%%%%%%%%%%%%%%%%%%%%%%%%%%%%%%%%%%%%%%%%%%%%%%%%%%%%%%%%%%%%%%%%%%%%%%%%%%%%%%%%%%%%%%%%%%%%%%%%%%%%%%%%%%%%%%%%%%
\section{Coherent states}
%%%%%%%%%%%%%%%%%%%%%%%%%%%%%%%%%%%%%%%%%%%%%%%%%%%%%%%%%%%%%%%%%%%%%%%%%%%%%%%%%%%%%%%%%%%%%%%%%%%%%%%%%%%%%%%%%%%%%%%%%%%%%%%%%%%%%%%%

In accordance with the algebraic approach developed until now, let us construct in this section a class of Barut-Girardello coherent states
\cite{Barut} related to the RS functions which are eigenstates of the lowering operator $\bop$ with eigenvalues $\mu \in \mathbb{C}$, that is,
\be
\lb{e31}
\bop \mathscr{F}_{\mu}(z;q) = \mu \mathscr{F}_{\mu}(z;q) . 
\ee
Within this context, the complex function $\mathscr{F}_{\mu}(z;q)$ can be properly expanded in terms of the complete set 
$\{ \Psi_{n}^{\inrs}(z;q) \}_{n \in \mathbb{N}}$ as follows \cite{Arik}:
\be
\lb{e32}
\mathscr{F}_{\mu}(z;q) = \sum_{n \in \mathbb{N}} \mathscr{C}_{n}(\mu;q) \Psi_{n}^{\inrs}(z;q) .
\ee
The coefficients $\{ \mathscr{C}_{n}(\mu;q) \}_{n \in \mathbb{N}^{\ast}}$ are then determined by means of equation (\ref{e31}), while the 
coefficient $\mathscr{C}_{0}(\mu;q)$ results from the orthogonality relation (\ref{e16}). Consequently, such mathematical procedure leads us 
to obtain 
\bd
\mathscr{C}_{n}(\mu;q) = e_{q}^{-1/2} \lpar (1-q) | \mu |^{2} \rpar \frac{\mu^{n}}{\sqrt{[n]_{q}!}} ,
\ed
where $e_{q}(x) \coloneq (x;q)_{\infty}^{-1}$ defines a $q$-exponential function which converges absolutely for $q \in (0,1)$ and $|x| < 1$ 
\cite[page 9]{Gasper}. Therefore, if one substitutes these coefficients in the right-hand side of equation (\ref{e32}) and takes into account 
the identity
\bd
\fl \qquad \quad \sum_{n \in \mathbb{N}} \mathscr{H}_{n}(z;q) \, \frac{\lbk \sqrt{q(1-q)} \, \mu \rbk^{n}}{(q;q)_{n}} = \lpar 
\sqrt{q(1-q)} \, \mu, \sqrt{q(1-q)} \, \mu z; q \rpar_{\infty}^{-1} ,
\ed
which, as it has been pointed out in section 2, can be promptly attained via equation (\ref{e2}), we finally obtain the entire function
\be
\lb{e33}
\mathscr{F}_{\mu}(z;q) = \frac{e_{q}^{-1/2} \lpar (1-q) | \mu |^{2} \rpar \mathscr{M}(z;q)}{\sqrt{2 \pi} \lpar \sqrt{q(1-q)} \, 
\mu, \sqrt{q(1-q)} \, \mu z; q \rpar_{\infty}} .
\ee
Note that equations (\ref{e32}) and (\ref{e33}) represent two different forms of expressing $\mathscr{F}_{\mu}(z;q)$, and that
$| \mathscr{C}_{n}(\mu;q) |^{2}$ is associated with the excitation probability distribution for the $q$-deformed coherent state. 
In the following, let us present at least two inherent algebraic properties of these particular coherent states. 

The first property is a direct consequence of expansion (\ref{e33}) and corresponds to the formula for the scalar product
\be
\lb{e34}
\fl \int_{-\pi}^{\pi} \mathscr{F}_{\nu}^{\ast} \Bigl(- q^{- \half} e^{\nc \varphi};q \Bigr) \mathscr{F}_{\mu} \Bigl(- q^{- \half} 
e^{\nc \varphi};q \Bigr) d \varphi = \frac{e_{q} \lpar (1-q) \mu \nu^{\ast} \rpar}{\lbk e_{q} \lpar (1-q) | \mu |^{2} \rpar e_{q} 
\lpar (1-q) | \nu |^{2} \rpar \rbk^{\half}} ,
\ee
from which we can infer the inequality
\bd
0 < \frac{\left| e_{q} \lpar (1-q) \mu \nu^{\ast} \rpar \right|^{2}}{e_{q} \lpar (1-q) | \mu |^{2} \rpar e_{q} \lpar (1-q) | \nu |^{2} 
\rpar} \leq 1 .
\ed
Note that this overlap probability is equal to one for $\nu = \mu$ (normalizability condition of the scalar product) and falls to zero 
when $| \mu - \nu |^{2}$ becomes large in the limit $q \rightarrow 1^{-}$; in other words, the coherent states $\{ \mathscr{F}_{\mu}(z;q)
\}_{\mu \in \mathbb{C}}$ are not orthogonal. Gray and Nelson \cite{Gray} obtained an analogous mathematical result for the scalar product 
of coherent states associated with the $q$-deformed HO, whose commutation relations are characterized by: ${\bf A} {\bf A}^{\dagger} - q^{- 1/2} 
{\bf A}^{\dagger} {\bf A} = q^{- {\bf N}/2}$, $[ {\bf N},{\bf A}^{\dagger} ] = {\bf A}^{\dagger}$, and $[ {\bf N},{\bf A} ] = - {\bf A}$. Such 
result is not a mere coincidence since both approaches can be considered as particular cases of the generalized $q$-deformed Heisenberg-Weyl 
algebra $\mathrm{U}_{q}^{(\alf,\bet,\gam)}(\mathfrak{h}_{4})$ \cite{Marchiolli}.

The next property assures the implicit resolution of unity (or completeness relation) for the coherent states defined by means of the 
entire function $\mathscr{F}_{\mu}(z;q)$, namely
\be
\lb{e35}
\int_{\mathbb{D}_{q}} \mathscr{F}_{\mu}^{\ast}(w;q) \mathscr{F}_{\mu}(z;q) \, d^{2} \sigma_{q}(\mu) = \lim_{\eps \rightarrow 1^{-}} 
\mathrm{K}_{\eps}(w,z;q) ,
\ee
where
\bd
\mathbb{D}_{q} = \lbr \mu = | \mu | e^{\nc \th} : | \mu |^{2} \in \Bigl[ 0, \frac{1}{q-1} \Bigr) \; \mbox{and} \; \th \in [0,2\pi) \rbr
\ed
corresponds to the integration domain in the complex plane,
\be
\lb{e36}
d^{2} \sigma_{q}(\mu) = \frac{e_{q} \lpar (1-q) | \mu |^{2} \rpar}{e_{q} \lpar (1-q)q | \mu |^{2} \rpar} \frac{d_{q} ( | \mu |^{2} ) 
d \th}{2 \pi}   
\ee
its respective measure \cite{Perelomov}, and $\mathrm{K}_{\eps}(w,z;q)$ represents the bilinear kernel (\ref{e17}). To demonstrate this 
specific identity, we first substitute the expansion (\ref{e32}) into the left-hand side of equation (\ref{e35}) and then carry out, 
subsequently, the integration over the angle variable $\th$. Such usual procedure permits to reduce the integration over the domain 
$\mathbb{D}_{q}$ in the Jackson $q$-integral \cite{Arik,Jackson2}
\bd
\fl \qquad \mathpzc{I}_{q}(n) = \int_{0}^{\case{1}{1-q}} \frac{| \mu |^{2n}}{e_{q} \lpar (1-q)q | \mu |^{2} \rpar} \, d_{q} ( | \mu |^{2} ) =
\frac{1}{(1-q)^{n}} \sum_{k \in \mathbb{N}} \frac{q^{k(n+1)}}{e_{q} \lpar q^{k+1} \rpar} .
\ed
As an intermediate stage in our proof, let us now consider the result derived in \cite{Gasper} for the $q$-exponential function $e_{q} \lpar 
q^{k+1} \rpar$, \ie, $e_{q}^{-1} \lpar q^{k+1} \rpar = (q;q)_{\infty} / (q;q)_{k}$. Consequently, substituting this result into 
$\mathpzc{I}_{q}(n)$, it is then immediate to show that
\bd
\fl \qquad \mathpzc{I}_{q}(n) = \frac{(q;q)_{\infty}}{(1-q)^{n}} \sum_{k \in \mathbb{N}} \frac{q^{k(n+1)}}{(q;q)_{k}} = 
\frac{(q;q)_{\infty}}{(1-q)^{n} (q^{n+1};q)_{\infty}} = \frac{(q;q)_{n}}{(1-q)^{n}} = [n]_{q}! . 
\ed
So, after some minor adjustments in our calculations, the right-hand side of equation (\ref{e35}) can be promptly reached. \ding{114}

To finish this section, let us discuss three important points raised by the identity (\ref{e35}). The first point corresponds to the resolution 
of unity for the coherent states which is implicitly demonstrated in our evaluations. By its turn, combining this result with the first
property, we can conclude that equation (\ref{e31}) produces a special set of overcomplete $q$-deformed coherent states. The second point is
entirely related to the term appearing in the right-hand side of equation (\ref{e35}): it reflects the normalization condition for the complex
representations used in this work. Finally, the last point is associated with the evidence that $d^{2} \sigma_{q}(\mu)$ may not be unique as 
some studies that appeared in the literature \cite{Borzov,Quesne} indicate. In fact, different integration measures can produce distinct 
integrals $\mathpzc{I}_{q}(n)$ whose integrands, by their turn, are related to solutions of Stieltjes and Hausdorff moment problems
\cite{Akhiezer,Klauder}.

%%%%%%%%%%%%%%%%%%%%%%%%%%%%%%%%%%%%%%%%%%%%%%%%%%%%%%%%%%%%%%%%%%%%%%%%%%%%%%%%%%%%%%%%%%%%%%%%%%%%%%%%%%%%%%%%%%%%%%%%%%%%%%%%%%%%%%%%
\section{Phase states}
%%%%%%%%%%%%%%%%%%%%%%%%%%%%%%%%%%%%%%%%%%%%%%%%%%%%%%%%%%%%%%%%%%%%%%%%%%%%%%%%%%%%%%%%%%%%%%%%%%%%%%%%%%%%%%%%%%%%%%%%%%%%%%%%%%%%%%%%

Previously, in section 3, we have established the $q$-differential forms for the lowering, raising and number operators whose respective
actions on the RS functions resemble those usually obtained for the annihilation, creation and number operators when acting on the 
wavefunctions $\{ \Psi_{n}^{\inho}(x;\alf) \}_{n \in \mathbb{N}}$ associated with the usual harmonic oscillator. This particular analogy 
leads us to investigate the possibility of constructing a polar decomposition for the lowering (raising) operator ${\bf B}$ 
$( {\bf B}^{\dagger} )$ through the equivalence relation ${\bf B} \equiv [ {\bf N} + {\bf 1} ]_{q}^{1/2} \opexp_{-}$ $( {\bf B}^{\dagger} 
\equiv \opexp_{+} [ {\bf N} + {\bf 1} ]_{q}^{1/2} )$, where ${\bf N}$ denotes the standard number operator.\footnote{It is important to mention 
that ${\bf N}_{q} \equiv {\bf B}^{\dagger} {\bf B}$ coincides with the standard number operator ${\bf N}$ only in the limit $q \rightarrow 1^{-}$ 
\cite{Marchiolli}. In this case, the operator ${\bf N}$ is subjected to the commutation relations $[ {\bf N},{\bf B} ] = - {\bf B}$ and 
$[ {\bf N},{\bf B}^{\dagger} ] = {\bf B}^{\dagger}$, which differ, by its turn, of those obtained in equation (\ref{e30}) for ${\bf N}_{q}$.} 
Moreover, $\opexp_{\mp}$ represent two `exponential' operators which will be explored adequately in this section. It is important to 
stress that the underlying problems of this specific decomposition and their possible solutions with convenient inherent mathematical 
properties were extensively discussed in the literature \cite{Review,Perina}. Here, our focus will be the construction of orthogonal
eigenstates related to the $q$-deformed cosine and sine operators, in analogy with the results discussed in \cite{Susskind,Carruthers},
namely
\be
\lb{e37}
\fl \quad \qquad \opcos \coloneq \frac{1}{2} \lpar \opexp_{-} + \opexp_{+} \rpar \quad \mbox{and} \quad \opsin \coloneq \frac{1}{2 \nc} 
\lpar \opexp_{-} - \opexp_{+} \rpar ,
\ee
as well as the calculation of certain mean values pertain to $\opcos$ and $\opsin$ which allow us to determine some intrinsic properties 
of the coherent states described in section 4.

Initially, let us assume that $\opexp_{\mp}$ acting on the RS functions $\Psi_{n}^{\inrs}(z;q)$ results in the identity $\opexp_{\mp}
\Psi_{n}^{\inrs}(z;q) \equiv \Psi_{n \mp 1}^{\inrs}(z;q)$, \ie, its action decreases/increases the excitation degree $n$ of the 
$q$-deformed HO by one \cite{Susskind}. The next step then consists in solving the eigenvalue equation
\be
\lb{e38}
\opcos \mathscr{X}_{\gamma}(z;q) = \cos (\gamma) \mathscr{X}_{\gamma}(z;q)
\ee
following the mathematical recipe described in \cite{Carruthers}. Note that $\{ \mathscr{X}_{\gamma}(z;q) \}_{\gamma \in [0,\pi]}$ can be 
expanded in terms of the complete set $\{ \Psi_{n}^{\inrs}(z;q) \}_{n \in \mathbb{N}}$, and their respective coefficients determined in
order to obey the eigenvalue equation (\ref{e38}). Thus, after some nontrivial algebra, we obtain
\be
\lb{e39}
\mathscr{X}_{\gamma}(z;q) = \sqrt{\frac{2}{\pi}} \, \sum_{n \in \mathbb{N}} \sin [ (n+1) \gamma ] \Psi_{n}^{\inrs}(z;q) ,
\ee
which satisfies not only the orthogonality relation
\be
\lb{e40}
\fl \qquad \quad \int_{- \pi}^{\pi} \mathscr{X}_{\gamma^{\prime}}^{\ast} \Bigl(- q^{- \half} e^{\nc \varphi};q \Bigr) \mathscr{X}_{\gamma} 
\Bigl(- q^{- \half} e^{\nc \varphi};q \Bigr) d \varphi = \delta (\gamma^{\prime} - \gamma)
\ee
but also the resolution of unity implicitly expressed as
\be
\lb{e41}
\int_{0}^{\pi} \mathscr{X}_{\gamma}^{\ast}(w;q) \mathscr{X}_{\gamma}(z;q) \, d \gamma = \lim_{\eps \rightarrow 1^{-}} \mathrm{K}_{\eps}
(w,z;q) .
\ee
Now, let us mention an important feature inherent to expansion (\ref{e39}): it vanishes at the points $\gamma = 0$ or $\pi$, and this fact 
implies the non-existence of singularities at these points since there are no states related to them \cite{Carruthers}.  

The construction process of the eigenfunctions related to $\opsin$ follows along similar lines, namely, they are solutions of the eigenvalue
equation
\be
\lb{e42}
\opsin \mathscr{Y}_{\gamma}(z;q) = \sin (\gamma) \mathscr{Y}_{\gamma}(z;q)
\ee
whose expansion in terms of the RS functions obeys the equation
\be
\lb{e43}
\fl \qquad \quad \mathscr{Y}_{\gamma}(z;q) = \nc \sqrt{\frac{2}{\pi}} \, \sum_{n \in \mathbb{N}} e^{\nc (n+1) \case{\pi}{2}} \sin \lbk 
(n+1) \lpar \gamma - \frac{\pi}{2} \rpar \rbk \Psi_{n}^{\inrs}(z;q) .
\ee
In analogy with properties (\ref{e40}) and (\ref{e41}) we also find the following relations:
\brr
\lb{e44}
\fl \qquad \quad \int_{- \pi}^{\pi} \mathscr{Y}_{\gamma^{\prime}}^{\ast} \Bigl(- q^{- \half} e^{\nc \varphi};q \Bigr) \mathscr{Y}_{\gamma} 
\Bigl(- q^{- \half} e^{\nc \varphi};q \Bigr) d \varphi = \delta (\gamma^{\prime} - \gamma) , \\
\lb{e45}
\fl \qquad \quad \int_{-\case{\pi}{2}}^{\case{\pi}{2}} \mathscr{Y}_{\gamma}^{\ast}(w;q) \mathscr{Y}_{\gamma}(z;q) \, d \gamma = 
\lim_{\eps \rightarrow 1^{-}} \mathrm{K}_{\eps}(w,z;q) .
\err
It is worth stressing that the eigenfunctions here obtained for the Hermitian operators $\opcos$ and $\opsin$ depend strongly on the
action of $\opexp_{\mp}$ upon the complete set $\{ \Psi_{n}^{\inrs}(z;q) \}_{n \in \mathbb{N}}$. In other words, different polar 
decompositions for the lowering operator (as well as distinct assumptions about the action of $\opexp_{\mp}$) lead us to derive different 
expressions for $\mathscr{X}_{\gamma}(z;q)$ and $\mathscr{Y}_{\gamma}(z;q)$. This fact was properly explored by Bergou and Englert 
\cite{Bergou} within the Wigner function context and its asymptotic form for a quantum operator, where, in particular, the authors 
showed how to construct different phase operators related to the usual HO.

%%%%%%%%%%%%%%%%%%%%%%%%%%%%%%%%%%%%%%%%%%%%%%%%%%%%%%%%%%%%%%%%%%%%%%%%%%%%%%%%%%%%%%%%%%%%%%%%%%%%%%%%%%%%%%%%%%%%%%%%%%%%%%%%%%%%%%%%%%
\section{Applications}
%%%%%%%%%%%%%%%%%%%%%%%%%%%%%%%%%%%%%%%%%%%%%%%%%%%%%%%%%%%%%%%%%%%%%%%%%%%%%%%%%%%%%%%%%%%%%%%%%%%%%%%%%%%%%%%%%%%%%%%%%%%%%%%%%%%%%%%%%%
\subsection{Mean values}
%%%%%%%%%%%%%%%%%%%%%%%%%%%%%%%%%%%%%%%%%%%%%%%%%%%%%%%%%%%%%%%%%%%%%%%%%%%%%%%%%%%%%%%%%%%%%%%%%%%%%%%%%%%%%%%%%%%%%%%%%%%%%%%%%%%%%%%%%%

In the following, we derive a set of closed-form expressions for certain moments related to the cosine and sine operators evaluated via 
coherent states $\{ \mathscr{F}_{\mu}(z;q) \}_{\mu \in \mathbb{C}}$. Our main task then consists in computing initially some specific 
mean values involving $\opcos$ and $\opsin$ by means of the auxiliary relation
\be
\lb{e46}
\lg {\bf O} \rg_{\mu} = \int_{-\pi}^{\pi} \mathscr{F}_{\mu}^{\ast} \Bigl(- q^{- \half} e^{\nc \varphi};q \Bigr) {\bf O} \mathscr{F}_{\mu} 
\Bigl(- q^{- \half} e^{\nc \varphi};q \Bigr) d \varphi .
\ee
It is important to mention that the action of such operators on the coherent states is not trivial, namely it presents peculiarities
inherent to the nonunitarity of $\opexp_{-}$. So, after lengthy calculations, we obtain for $\mu = | \mu | e^{\nc \th}$ the exact 
results \cite{Nelson}
\brr
\fl \quad \lg \opcos \rg_{\mu} = | \mu | \, e_{q}^{-1} \lpar (1-q) | \mu |^{2} \rpar \mathpzc{M}_{q}(| \mu |^{2}) \cos (\th) , \nn \\
\fl \quad \lg \opsin \rg_{\mu} = | \mu | \, e_{q}^{-1} \lpar (1-q) | \mu |^{2} \rpar \mathpzc{M}_{q}(| \mu |^{2}) \sin (\th) , \nn \\
\fl \quad \lg \bcal{C}^{2} \rg_{\mu} = \frac{1}{2} - \frac{1}{4} e_{q}^{-1} \lpar (1-q) | \mu |^{2} \rpar + \frac{1}{2} | \mu |^{2} 
e_{q}^{-1} \lpar (1-q) | \mu |^{2} \rpar \mathpzc{N}_{q}(| \mu |^{2}) \cos (2 \th) , \nn \\
\fl \quad \lg \bcal{S}^{2} \rg_{\mu} = \frac{1}{2} - \frac{1}{4} e_{q}^{-1} \lpar (1-q) | \mu |^{2} \rpar - \frac{1}{2} | \mu |^{2} 
e_{q}^{-1} \lpar (1-q) | \mu |^{2} \rpar \mathpzc{N}_{q}(| \mu |^{2}) \cos (2 \th) , \nn \\
\fl \quad \lg \bcal{C}^{2} + \bcal{S}^{2} \rg_{\mu} = 1 - \frac{1}{2} e_{q}^{-1} \lpar (1-q) | \mu |^{2} \rpar , \nn \\
\fl \quad \lg \bcal{C}^{2} - \bcal{S}^{2} \rg_{\mu} = | \mu |^{2} e_{q}^{-1} \lpar (1-q) | \mu |^{2} \rpar \mathpzc{N}_{q}(| \mu |^{2}) 
\cos (2 \th) , \nn \\
\fl \quad \lg \opcos \opsin + \opsin \opcos \rg_{\mu} = | \mu |^{2} e_{q}^{-1} \lpar (1-q) | \mu |^{2} \rpar \mathpzc{N}_{q}(| \mu |^{2}) 
\sin (2 \th) , \nn \\
\fl \quad \lg \opcos \opsin - \opsin \opcos \rg_{\mu} = \frac{\nc}{2} e_{q}^{-1} \lpar (1-q) | \mu |^{2} \rpar , \nn
\err
where the functions $\mathpzc{M}_{q}(| \mu |^{2})$ and $\mathpzc{N}_{q}(| \mu |^{2})$ are here defined by the power series
\bd
\fl \quad \mathpzc{M}_{q}(| \mu |^{2}) = \sum_{n \in \mathbb{N}} \frac{| \mu |^{2n}}{[ n ]_{q}! \lpar [ n+1 ]_{q} \rpar^{\half}} \quad 
\mbox{and} \quad \mathpzc{N}_{q}(| \mu |^{2}) = \sum_{n \in \mathbb{N}} \frac{| \mu |^{2n}}{[ n ]_{q}! \lpar [ n+2 ]_{q} [ n+1 ]_{q}
\rpar^{\half}} .
\ed
Such mean values can be interpreted as the $q$-deformed version of those obtained by Carruthers and Nieto \cite{Carruthers}. Furthermore, 
for $q \rightarrow 1^{-}$ and $| \mu |^{2} \gg 1$, these exact results reach the asymptotic limits (\eg, see Lynch \cite[page 378]{Review}
for further discussion)
\brr
\fl & \; \lg \opcos \rg_{\mu} \approx \cos (\th) , \; \lg \opsin \rg_{\mu} \approx \sin (\th) , \; \lg \bcal{C}^{2} \rg_{\mu} \approx 
\cos^{2} (\th) , \; \lg \bcal{S}^{2} \rg_{\mu} \approx \sin^{2} (\th) , \; \lg \bcal{C}^{2} + \bcal{S}^{2} \rg_{\mu} \approx 1 , \nn \\
\fl & \; \lg \bcal{C}^{2} - \bcal{S}^{2} \rg_{\mu} \approx \cos (2 \th) , \; \lg \opcos \opsin + \opsin \opcos \rg_{\mu} \approx 
\sin (2 \th) , \; \lg \opcos \opsin - \opsin \opcos \rg_{\mu} \approx 0 . \nn
\err
Figure \ref{Mean Values} shows the plots of the first six mean values versus $| \mu |^{2} \in [0,5]$ with $\th = \case{\pi}{3}$ fixed,
and different values of $q$ such that $(1-q) | \mu |^{2} < 1$ be satisfied. The asymptotic limits observed in the numerical calculations
are in agreement with those predicted theoretically, and this fact will be our reference in the study of Robertson-Schr\"{o}dinger
uncertainty relations for the cosine and sine operators.  
%%%%%%%%%%%%%%%%%%%%%%%%%%%%%%%%%%%%%%%%%%%%%%%%%%%%%%%%%%%%%%%%%%%%%%%%%%%%%%%%%%%%%%%%%%%%%%%%%%%%%%%%%%%%%%%%%%%%%%%%%%%%%%%%%%%%%%%%%
\begin{figure}[!t]
\centering
\begin{minipage}[b]{0.45\linewidth}
\includegraphics[width=\textwidth]{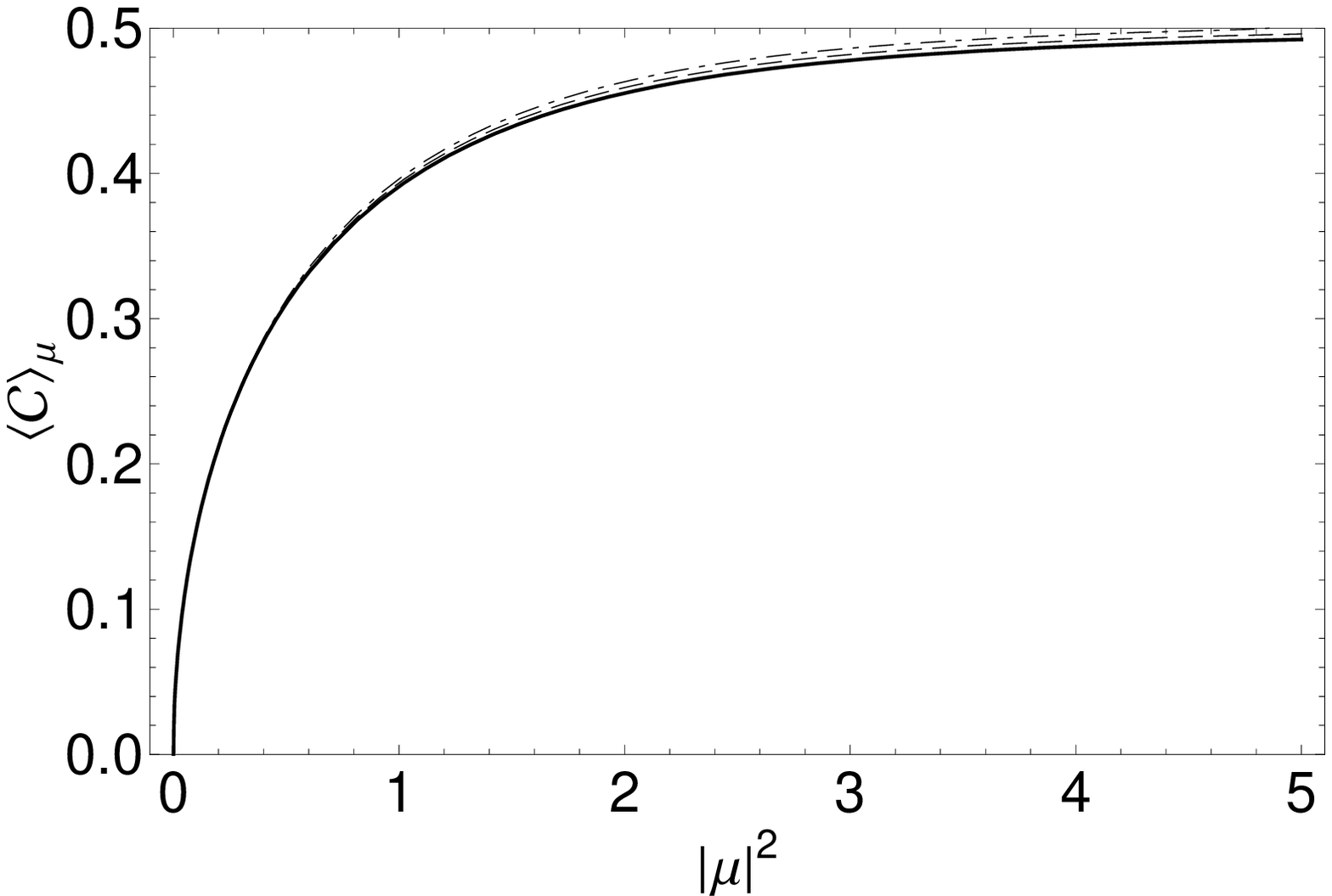}
\end{minipage} \hfill
\begin{minipage}[b]{0.45\linewidth}
\includegraphics[width=\textwidth]{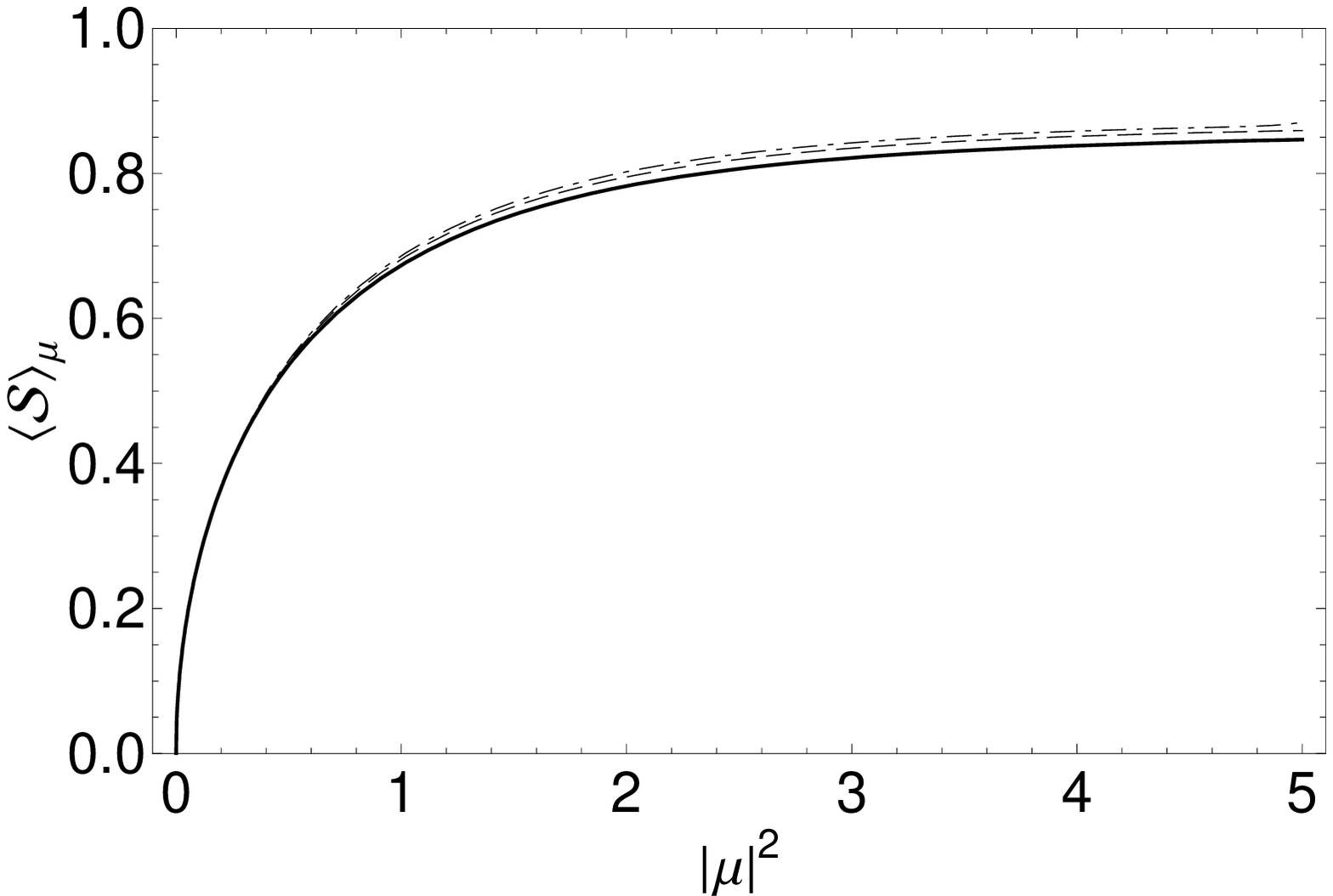}
\end{minipage} \hfill
\begin{minipage}[b]{0.45\linewidth}
\includegraphics[width=\textwidth]{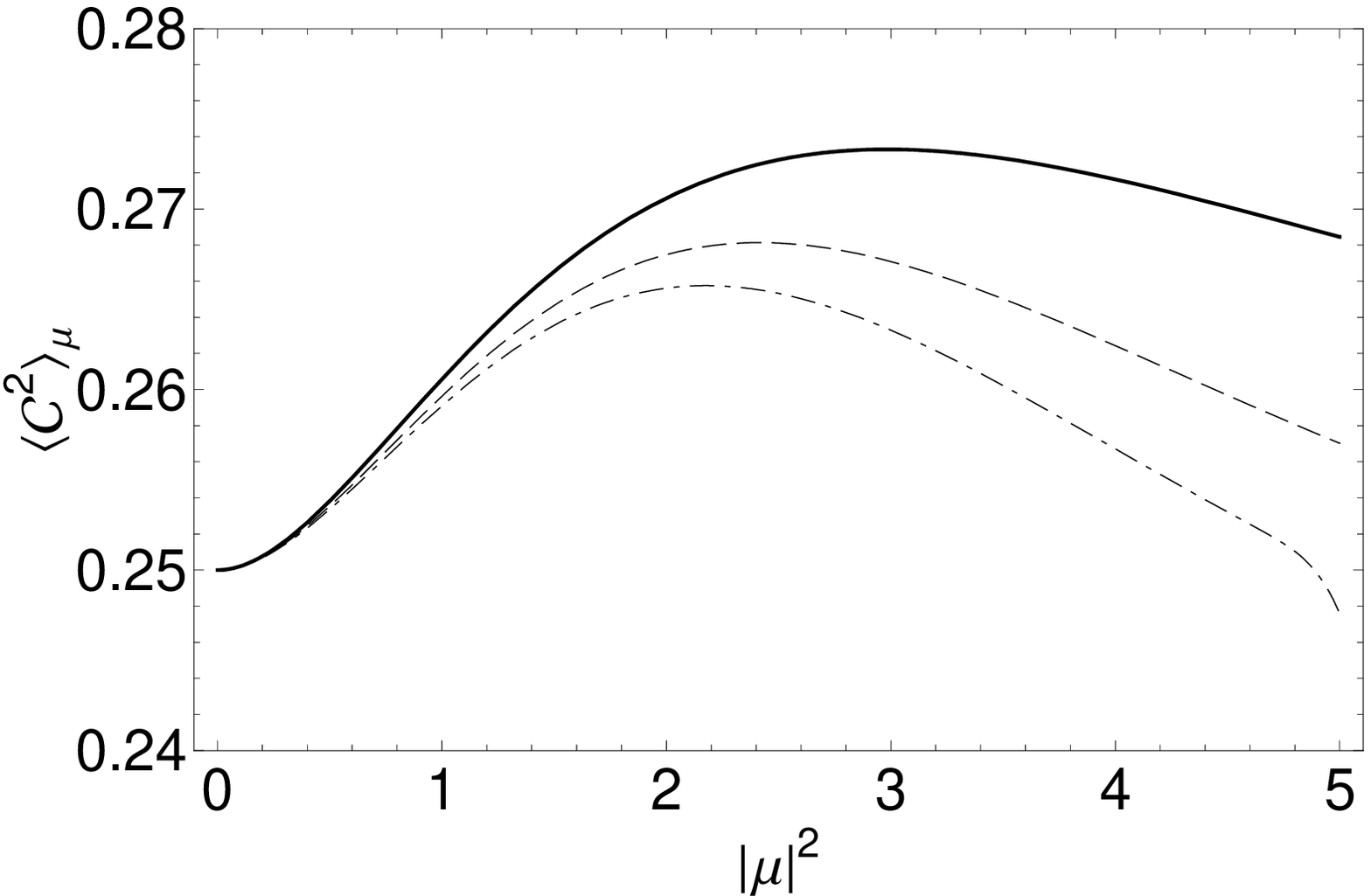}
\end{minipage} \hfill
\begin{minipage}[b]{0.45\linewidth}
\includegraphics[width=\textwidth]{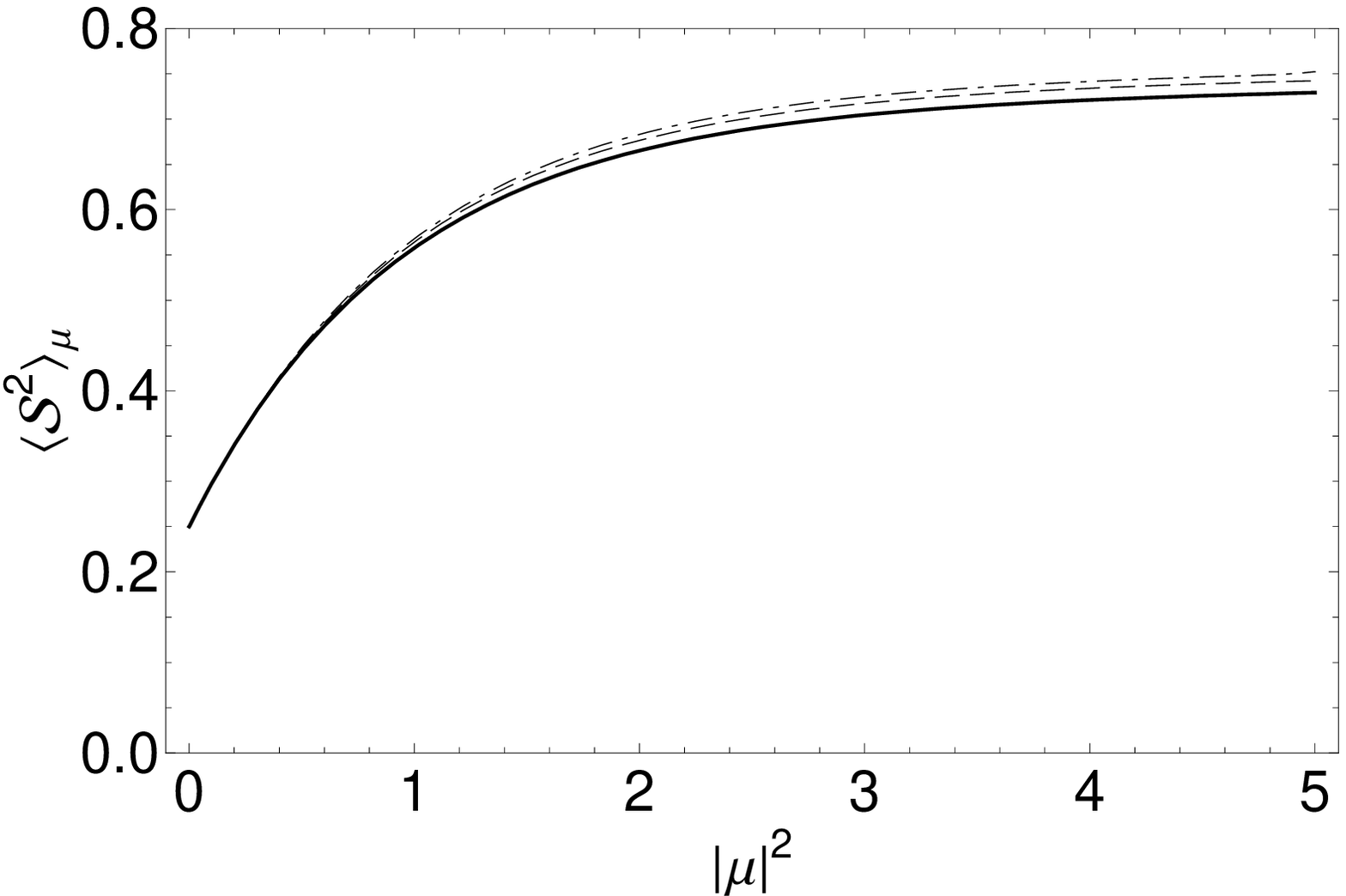}
\end{minipage} \hfill
\begin{minipage}[b]{0.45\linewidth}
\includegraphics[width=\textwidth]{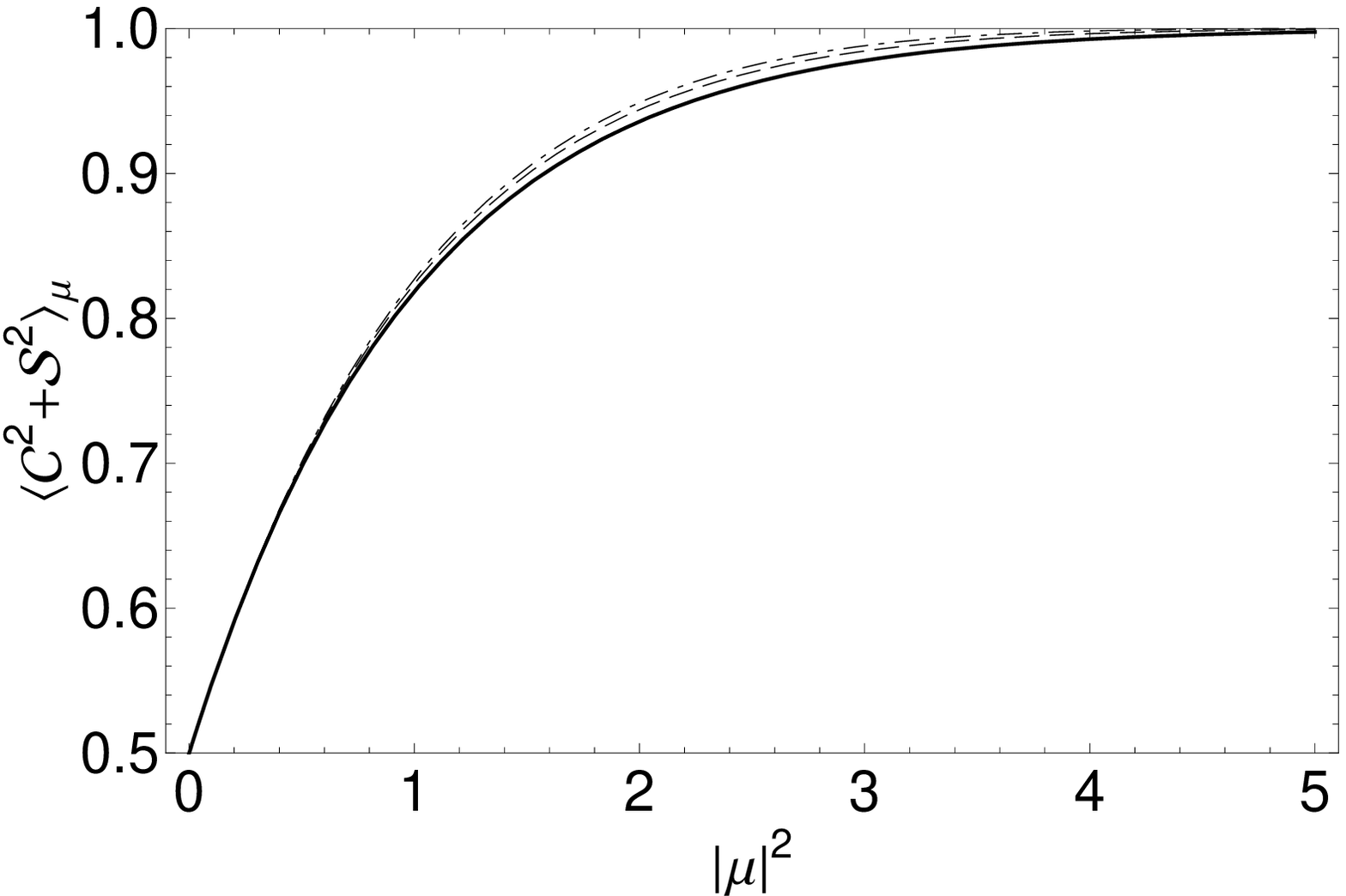}
\end{minipage} \hfill
\begin{minipage}[b]{0.45\linewidth}
\includegraphics[width=\textwidth]{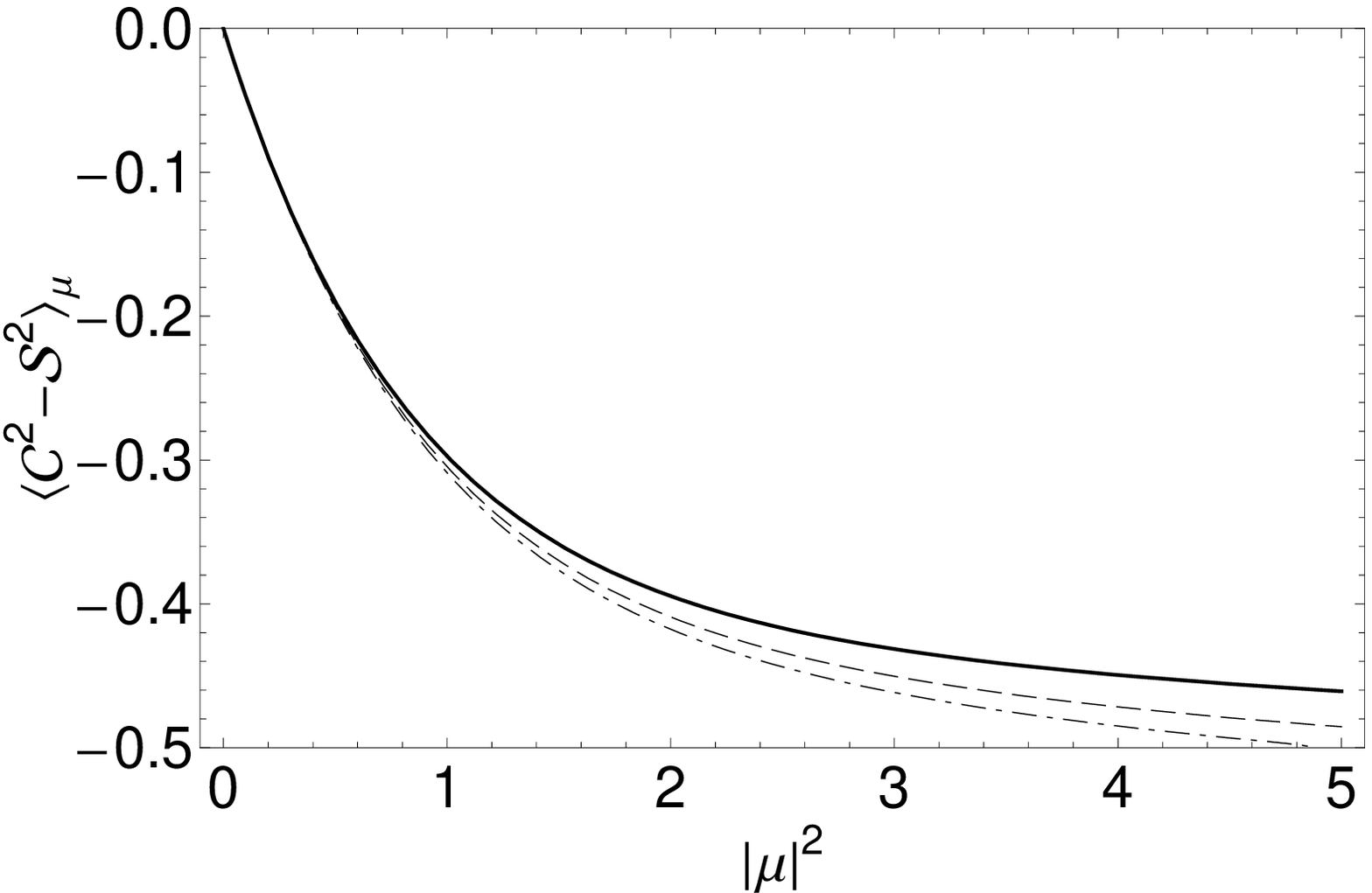}
\end{minipage}
\caption{\lb{Mean Values} Plots of first- and second-order moments involving the cosine and sine operators as a function of $0 \leq | 
\mu |^{2} \leq 5$ with $\th = \case{\pi}{3}$ fixed, and different values of $q$. In such examples, the dot-dashed, dashed, and solid 
lines correspond, respectively, to the values of $q = 0.8$, $0.85$, and $0.9$. Note that the distinct asymptotic values reached in each 
case exhibit a strong dependence on the parameters $q$ and $| \mu |^{2}$, this fact being associated with the convergence criterion
$(1-q) | \mu |^{2} < 1$ adopted for the $q$-exponential function.} 
\end{figure}
%%%%%%%%%%%%%%%%%%%%%%%%%%%%%%%%%%%%%%%%%%%%%%%%%%%%%%%%%%%%%%%%%%%%%%%%%%%%%%%%%%%%%%%%%%%%%%%%%%%%%%%%%%%%%%%%%%%%%%%%%%%%%%%%%%%%%%%%%%

%%%%%%%%%%%%%%%%%%%%%%%%%%%%%%%%%%%%%%%%%%%%%%%%%%%%%%%%%%%%%%%%%%%%%%%%%%%%%%%%%%%%%%%%%%%%%%%%%%%%%%%%%%%%%%%%%%%%%%%%%%%%%%%%%%%%%%%%%%
\subsection{Symmetrical uncertainty relation}
%%%%%%%%%%%%%%%%%%%%%%%%%%%%%%%%%%%%%%%%%%%%%%%%%%%%%%%%%%%%%%%%%%%%%%%%%%%%%%%%%%%%%%%%%%%%%%%%%%%%%%%%%%%%%%%%%%%%%%%%%%%%%%%%%%%%%%%%%%

As a last topic of interest, let us now study qualitatively the Robertson-Schr\"{o}dinger uncertainty relation \cite{Dodonov}
\be
\lb{e47}
\mathscr{U}_{\incs} \coloneq \mathscr{V}_{\inc} \mathscr{V}_{\ins} - ( \mathscr{V}_{\incs} )^{2} \geq \case{1}{4} | \lg [ \opcos,\opsin ] 
\rg_{\mu} |^{2}
\ee
through the mean values established for the coherent states, where
\bd
\fl \quad \mathscr{V}_{\inc} \equiv \lg \opcos^{2} \rg_{\mu} - \lg \opcos \rg_{\mu}^{2} , \; \; \mathscr{V}_{\ins} \equiv \lg \opsin^{2} 
\rg_{\mu} - \lg \opsin \rg_{\mu}^{2} , \; \; \mbox{and} \; \; \mathscr{V}_{\incs} \equiv \lg \half \{ \opcos,\opsin \} \rg_{\mu} - \lg
\opcos \rg_{\mu} \lg \opsin \rg_{\mu}
\ed
represent the variances related to the cosine and sine operators. In addition, the terms $\lg [ \opcos,\opsin ] \rg_{\mu}$ and
$\lg \{ \opcos,\opsin \} \rg_{\mu}$ correspond to the commutation and anticommutation relation mean values, respectively. Thus, after a 
straightforward calculation, we find that $\mathscr{U}_{\incs}$ does not depend on the angle variable $\th$, but only on the parameters $q$
and $| \mu |$. Indeed, using the results previously obtained in the last section, if one denotes
\brr
\mathfrak{a} &=& \frac{1}{2} - \frac{1}{4} e_{q}^{-1} \lpar (1-q) | \mu |^{2} \rpar , \nn \\
\mathfrak{b} &=& \frac{1}{2} | \mu |^{2} e_{q}^{-1} \lpar (1-q) | \mu |^{2} \rpar \mathpzc{N}_{q}(| \mu |^{2}) , \nn \\
\mathfrak{c} &=& | \mu | \, e_{q}^{-1} \lpar (1-q) | \mu |^{2} \rpar \mathpzc{M}_{q}(| \mu |^{2}) , \nn
\err
the left-hand side of equation (\ref{e47}) can be properly written as $\mathscr{U}_{\incs} = (\mathfrak{a} - \mathfrak{b})(\mathfrak{a} + 
\mathfrak{b} - \mathfrak{c}^{2})$. It is worth emphasizing that our numerical evaluations corroborate the inequality $\mathscr{U}_{\incs} 
\geq \case{1}{4} | \lg [ \opcos,\opsin ] \rg_{\mu} |^{2}$. Moreover, for $q \rightarrow 1^{-}$ and $| \mu |^{2} \gg 1$, we obtain $| \lg 
[ \opcos,\opsin ] \rg_{\mu} |^{2} \rightarrow 0$, which implies that $\opcos$ and $\opsin$ can be considered as commutative variables
\cite{Carruthers}. 

Next, let us derive a symmetrical relation which involves a particular combination of the number-cosine and number-sine Robertson-Schr\"{o}dinger
uncertainty relations, namely, 
\brr
\lb{e48}
& \mathscr{V}_{\inn} \mathscr{V}_{\inc} - ( \mathscr{V}_{\inn \inc} )^{2} \geq \case{1}{4} | \lg [ {\bf N},\opcos ] \rg_{\mu} |^{2} =
\case{1}{4} \lg \opsin \rg_{\mu}^{2} \\
\lb{e49}
& \mathscr{V}_{\inn} \mathscr{V}_{\ins} - ( \mathscr{V}_{\inn \ins} )^{2} \geq \case{1}{4} | \lg [ {\bf N},\opsin ] \rg_{\mu} |^{2} =
\case{1}{4} \lg \opcos \rg_{\mu}^{2}
\err
where $\mathscr{V}_{\inn} \equiv \lg {\bf N}^{2} \rg_{\mu} - \lg {\bf N} \rg_{\mu}^{2}$ represents the variance related to the nondeformed 
number operator, with
\bd
\fl \qquad \mathscr{V}_{\inn \inc} \equiv \lg \half \{ {\bf N},\opcos \} \rg_{\mu} - \lg {\bf N} \rg_{\mu} \lg \opcos \rg_{\mu} \quad
\mbox{and} \quad \mathscr{V}_{\inn \ins} \equiv \lg \half \{ {\bf N},\opsin \} \rg_{\mu} - \lg {\bf N} \rg_{\mu} \lg \opsin \rg_{\mu} 
\ed
being the corresponding covariances associated with the number-cosine and number-sine operators. Our next step consists in adding (\ref{e48}) 
and (\ref{e49}) with the aim of obtaining the symmetrical relation
\be
\lb{e50}
\mathscr{U}_{\mathrm{sym}} \equiv \frac{\mathscr{V}_{\inn} (\mathscr{V}_{\inc} + \mathscr{V}_{\ins}) - \lbk ( \mathscr{V}_{\inn \inc} )^{2} +
( \mathscr{V}_{\inn \ins} )^{2} \rbk}{\lg \opsin \rg_{\mu}^{2} + \lg \opcos \rg_{\mu}^{2}} \geq \frac{1}{4}
\ee
between $\opcos$ and $\opsin$, which does not depend on the angle variable $\th$. Indeed, in order to prove this assertion it is sufficient 
to calculate the additional mean values
\brr
\lg {\bf N}^{k} \rg_{\mu} = e_{q}^{-1} \lpar (1-q) | \mu |^{2} \rpar \sum_{n \in \mathbb{N}} \frac{n^{k} | \mu |^{2n}}{[ n ]_{q}!} \quad
(k \geq 0) , \nn \\
\lg \half \{ {\bf N},\opcos \} \rg_{\mu} = \frac{| \mu |}{2} e_{q}^{-1} \lpar (1-q) | \mu |^{2} \rpar \mathpzc{L}_{q}(| \mu |^{2})
\cos ( \th ) , \nn \\
\lg \half \{ {\bf N},\opsin \} \rg_{\mu} = \frac{| \mu |}{2} e_{q}^{-1} \lpar (1-q) | \mu |^{2} \rpar \mathpzc{L}_{q}(| \mu |^{2})
\sin ( \th ) , \nn
\err
with $\mathpzc{L}_{q}(| \mu |^{2})$ given by the power series
\bd
\mathpzc{L}_{q}(| \mu |^{2}) = \sum_{n \in \mathbb{N}} \frac{(2n+1) | \mu |^{2n}}{[ n ]_{q}! \lpar [ n+1 ]_{q} \rpar^{\half}} .
\ed
%
%%%%%%%%%%%%%%%%%%%%%%%%%%%%%%%%%%%%%%%%%%%%%%%%%%%%%%%%%%%%%%%%%%%%%%%%%%%%%%%%%%%%%%%%%%%%%%%%%%%%%%%%%%%%%%%%%%%%%%%%%%%%%%%%%%%%%%%%%%
\begin{figure}[!t]
\centering
\begin{minipage}[b]{0.45\linewidth}
\includegraphics[width=\textwidth]{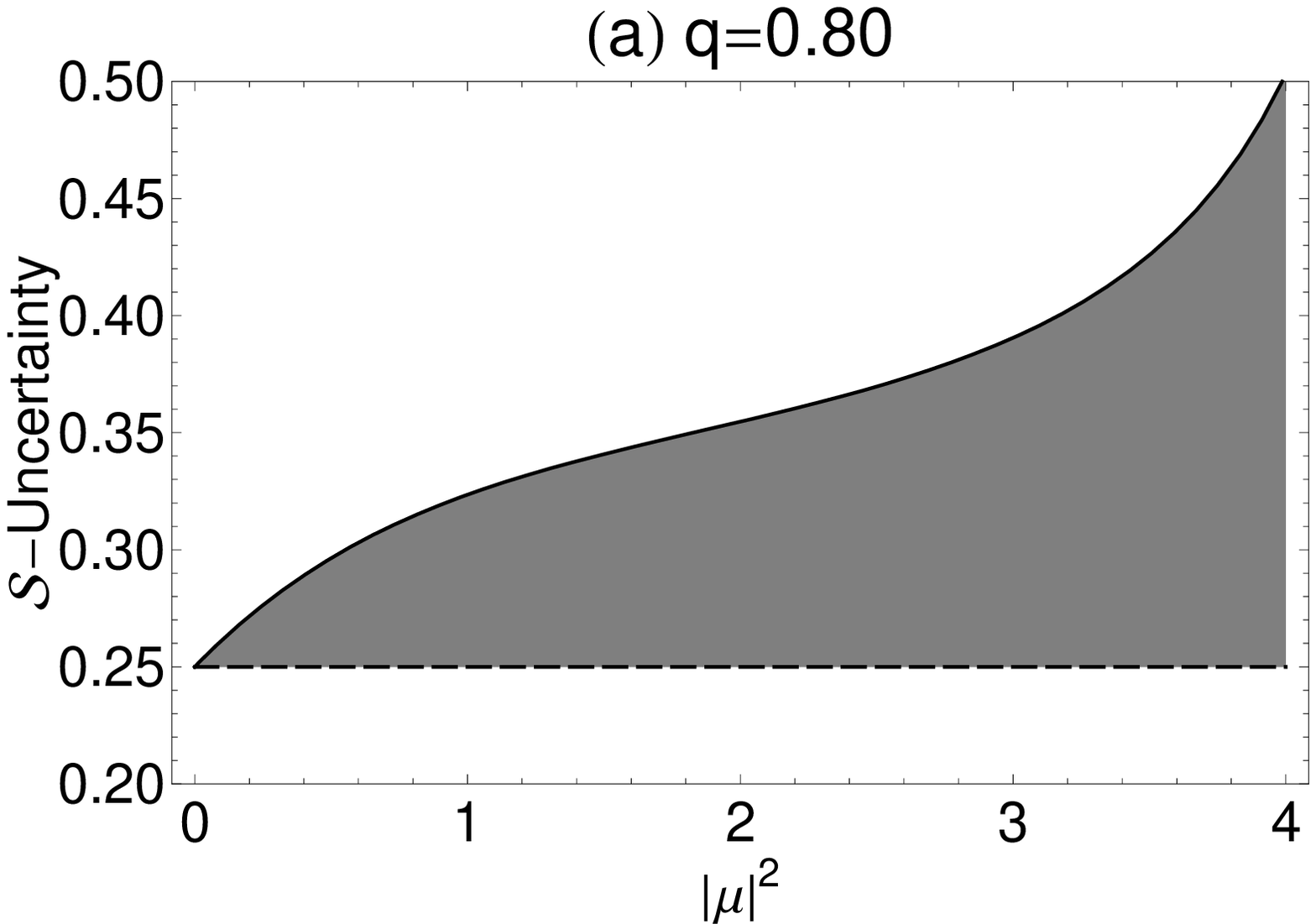}
\end{minipage} \hfill
\begin{minipage}[b]{0.45\linewidth}
\includegraphics[width=\textwidth]{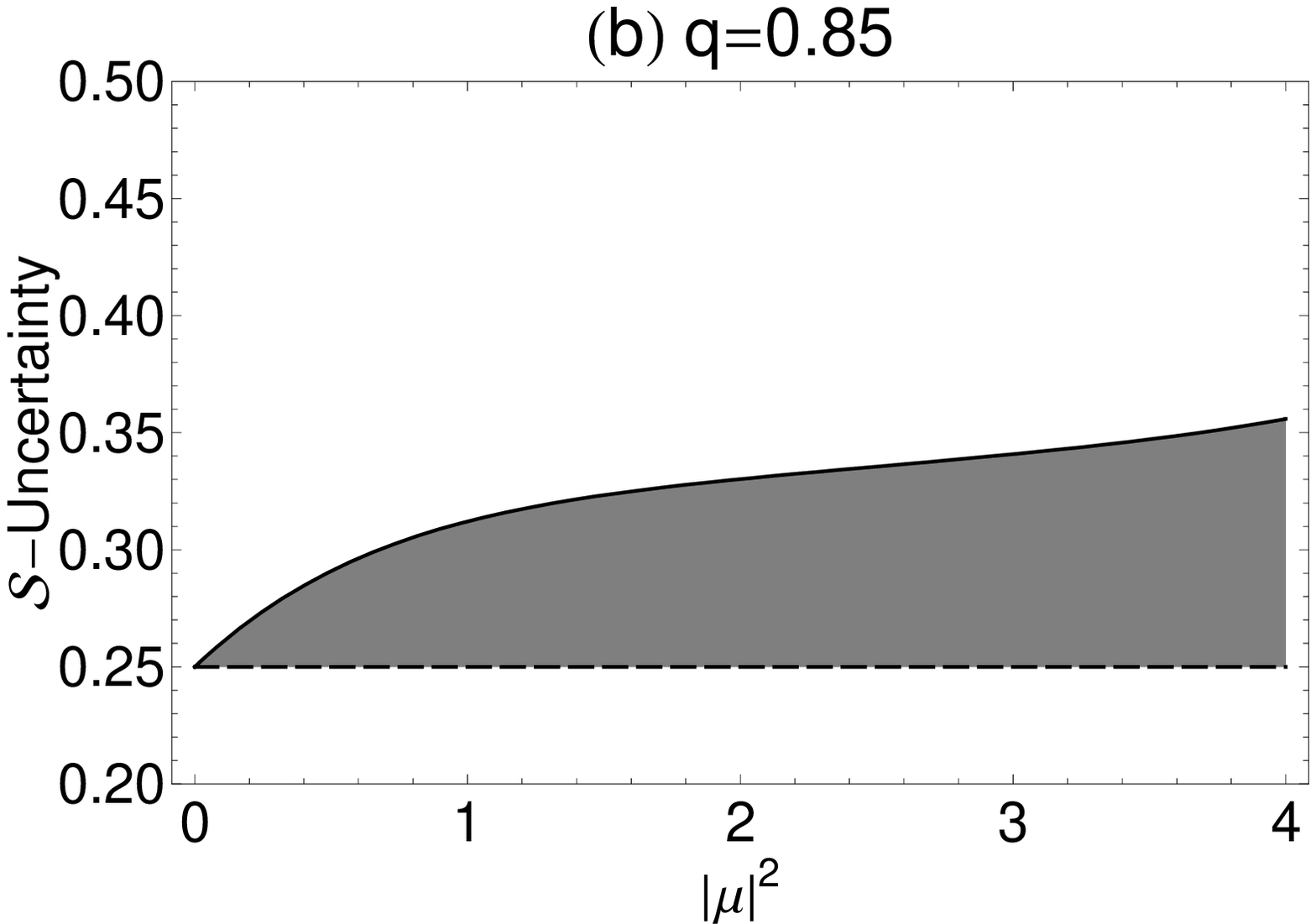}
\end{minipage} \hfill
\begin{minipage}[b]{0.45\linewidth}
\includegraphics[width=\textwidth]{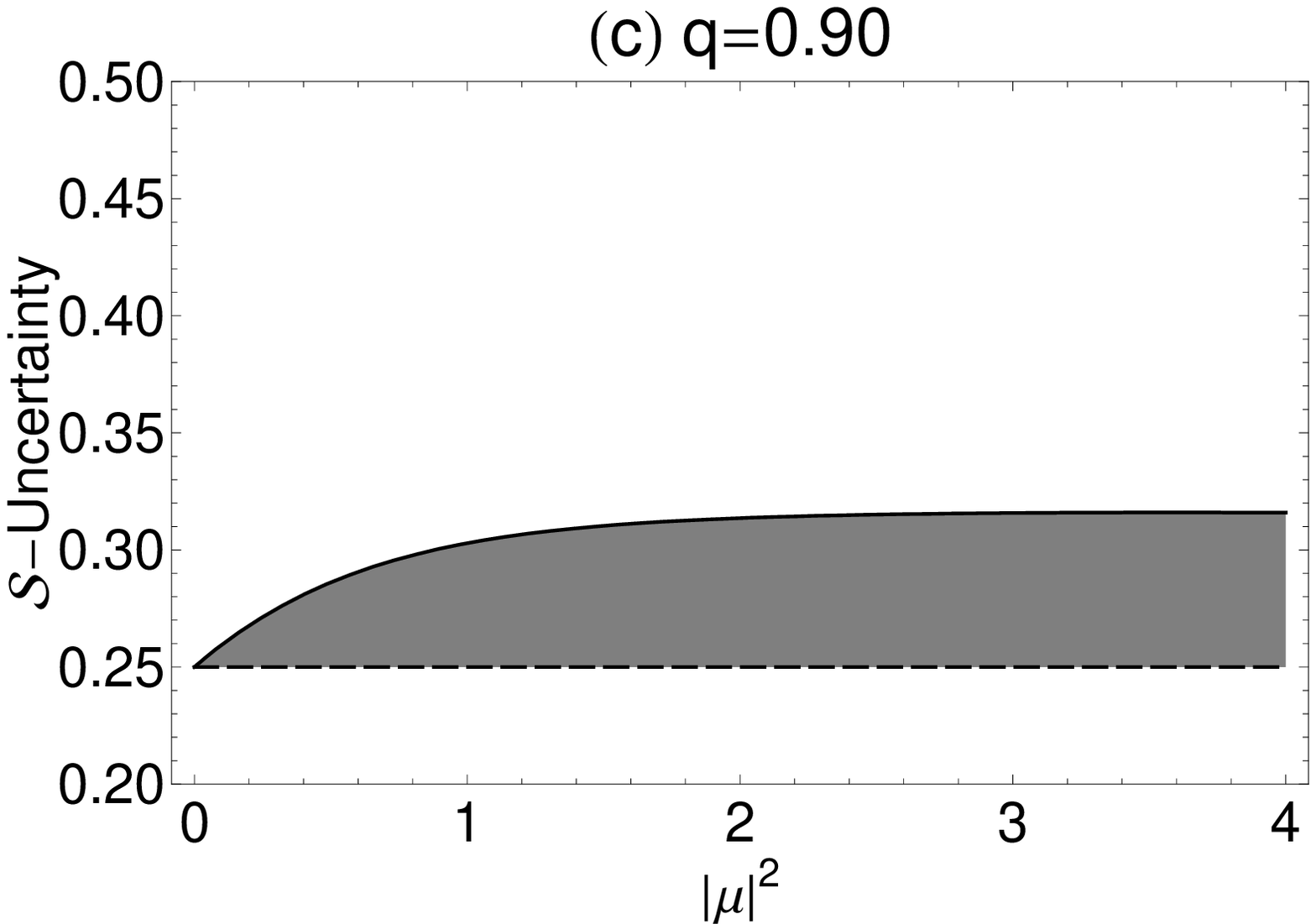}
\end{minipage} \hfill
\begin{minipage}[b]{0.45\linewidth}
\includegraphics[width=\textwidth]{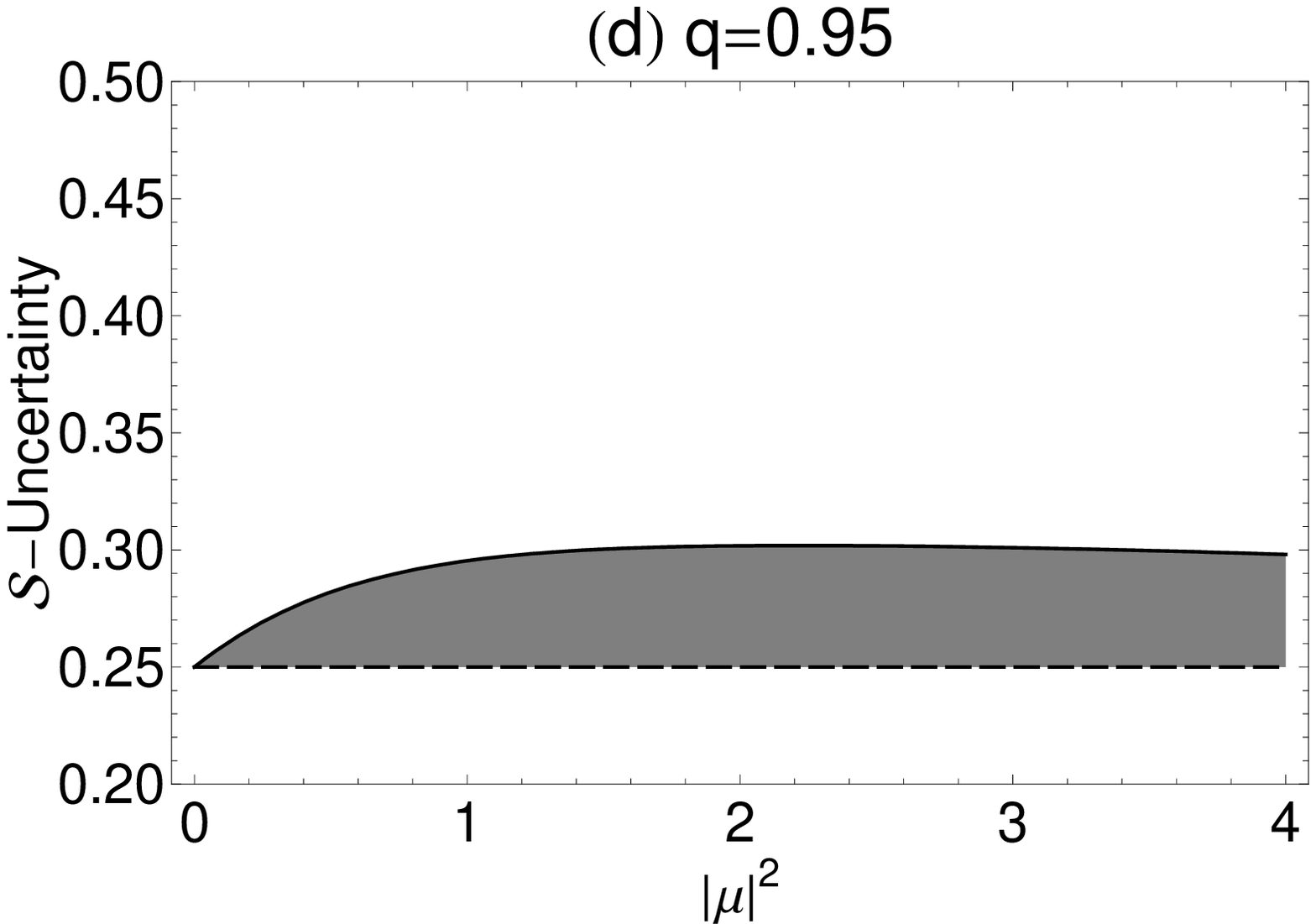}
\end{minipage}
\caption{\lb{S-Uncertainty} (Color online) Plots of $\mathscr{U}_{\mathrm{sym}}$ (solid line) as a function of $| \mu |^{2} \in [0,4]$ 
for different values of $q$. In all cases, the dashed line corresponds to the constant value $\case{1}{4}$, which reflects the pattern 
behaviour observed for the nondeformed coherent states. Note that for values of $q$ close to $1^{-}$, the space filled (gray color) 
between the curves diminishes, and when $| \mu |^{2} \gg 1$ we obtain the asymptotic limit $\mathscr{U}_{\mathrm{sym}} \rightarrow 
\case{1}{4}$.} 
\end{figure}
%%%%%%%%%%%%%%%%%%%%%%%%%%%%%%%%%%%%%%%%%%%%%%%%%%%%%%%%%%%%%%%%%%%%%%%%%%%%%%%%%%%%%%%%%%%%%%%%%%%%%%%%%%%%%%%%%%%%%%%%%%%%%%%%%%%%%%%%%%
Figure (\ref{S-Uncertainty}) illustrates the symmetrical relation (\ref{e50}) as a function of $| \mu |^{2} \in [0,4]$ for (a) $q=0.80$, (b) 
$q=0.85$, (c) $q=0.90$ and (d) $q=0.95$. In all these pictures, the solid line corresponds to the left-hand side of the inequality (\ref{e50}), 
while the right-hand side is represented by the dashed line. Now, let us say some few words about the space filled (gray color) between both
the curves: since the convergence criterion $0 \leq (1-q) | \mu |^{2} < 1$ restricts the domain of the variables $q$ and $| \mu |^{2}$, we can
verify that the area related to this space decreases for values of $q$ near to $1^{-}$ and $0 < | \mu |^{2} \leq 4$ --- see figures 4(a)-4(b).
A plausible explanation of this fact leads us to the $q$-deformed coherent states --- here represented by the eigenfunctions 
$\{ \mathscr{F}_{\mu}(z;q) \}_{\mu \in \mathbb{C}}$ --- and their inherent properties, \ie, these particular states are not minimum uncertainty
states (excepting the vacuum state $| \mu | = 0$), which justifies, in principle, the appearance of this specific region between the curves 
and its possible variation. As a last comment, let us mention that $\mathscr{V}_{\inn} \neq \lg {\bf N} \rg_{\mu}$ in such a case, which implies 
that $\{ | \mathscr{C}_{n}(\mu;q) |^{2} \}_{n \in \mathbb{N}}$ are not Poisson-distributed.

%%%%%%%%%%%%%%%%%%%%%%%%%%%%%%%%%%%%%%%%%%%%%%%%%%%%%%%%%%%%%%%%%%%%%%%%%%%%%%%%%%%%%%%%%%%%%%%%%%%%%%%%%%%%%%%%%%%%%%%%%%%%%%%%%%%%%%%%
\section{Concluding remarks}
%%%%%%%%%%%%%%%%%%%%%%%%%%%%%%%%%%%%%%%%%%%%%%%%%%%%%%%%%%%%%%%%%%%%%%%%%%%%%%%%%%%%%%%%%%%%%%%%%%%%%%%%%%%%%%%%%%%%%%%%%%%%%%%%%%%%%%%%

Within the scope of special functions in mathematics and physics, although the Rogers-Szeg\"{o} polynomials play an important role in
specific problems related to $q$-deformed algebras, they still remain practically unexplored if one considers the wide range of applications 
in quantum mechanics. A remarkable property of this sort of orthogonal polynomials states that $\{ \prz_{n}(z;q) \}_{n \in \mathbb{N}}$
are orthogonalized on the unit circle (pictorially represented in the complex plane) by means of a particular measure, the Jacobi
$\vartheta_{3}$-function \cite{Szego}. In the quantum-mechanical context, the benefits of this kind of {\it angular representation}, if 
employed in the fascinating problem of the polar decomposition of the annihilation operator concerning the usual harmonic oscillator 
(or, in other words, on the existence of a well-defined operator corresponding to the phase observable of the electromagnetic field in 
quantum optics), were not investigated until now or even mentioned in the literature. Here, we have presented an appreciable set of new 
interesting results which have potential applications not only in the investigative process on the different polar decompositions of the 
$q$-deformed annihilation operator \cite{Bergou}, where the aforementioned angular representation has an important rule, but also in the 
study of a $q$-analogue to the Jordan-Schwinger mapping for the angular momentum operators \cite{Marchiolli}.

Once the Szeg\"{o} measure can be decomposed in the complex plane, it is natural to construct a set of complex functions that not only
embodies such decomposition but also satisfies automatically an orthogonality relation analogous to that derived for the RS polynomials. 
Before we perform such task, some essential mathematical properties inherent to these polynomials have been adequately reviewed in our 
explanatory notes; in addition, we have also obtained two new integral representations for the RS polynomials which establish an important 
link with the $q$-Pochhammer symbol and the Stieltjes-Wigert polynomials \cite{Carlitz}. In fact, such results have paved the way for 
subsequent developments towards a solid framework of coherent and phase states conceived within a purely algebraic approach.

Named as Rogers-Szeg\"{o} functions and here denoted by $\{ \Psi_{n}^{\inrs}(z;q) \}_{n \in \mathbb{N}}$, our object of study was then
defined in terms of the product $\mathscr{R}_{n}(z;q) \mathscr{M}(z;q)$, where the first term $\mathscr{R}_{n}(z;q)$ presents a direct 
connection with the RS polynomials, while the second term $\mathscr{M}(z;q)$ represents a complex weight function. The particular
parametrization $z = - q^{- \half} e^{\nc \varphi}$ with $\varphi \in [-\pi,\pi]$ has allowed us to verify that $\Psi_{n}^{\inrs}(z;q)$ 
can be orthogonalized on a unit circle for any $q \in (0,1)$. Furthermore, we have also established a set of interesting formal results that
characterize its inherent algebraic properties. For example, the discussion about the completeness relation pertaining to the RS functions 
was carried out by means of the bilinear kernel $\mathrm{K}_{\eps}(w,z;q)$, which obeys certain properties that lead us to reveal the 
normalization condition for the complex representations used in this work to describe such functions. In the following, we have analyzed 
separately each term of $\Psi_{n}^{\inrs}(z;q)$ with the aim of obtaining not only new recurrence relations for $\mathscr{R}_{n}(z;q)$ 
but also some scaling relations for $\mathscr{M}(z;q)$ that involve odd and even powers of the parameter $q$. As a by-product of this 
analysis, we have introduced a specific definition of Jackson's $q$-derivative that permits us to determine two closed-form expressions which 
connect the action of $\mathpzc{D}_{q^{2}}$ over $\Psi_{n}^{\inrs}(z;q)$ with the different excitation degrees $n$ and $n \mp 1$, preserving, 
by its turn, the phase of the RS functions. So, the construction of $q$-differential forms of the lowering, raising and number operators from 
these results can be considered, at this stage, an immediate process. Those differential representations were then interpreted as the particular
realization of the IACK algebra which was characterized, within this context, through well-established commutation relations.

The applications of this algebraic approach in the construction process of coherent and phase states certainly represented an ideal scenario 
within our investigative theoretical framework. In this way, we adopted in a first moment the mathematical procedure developed by 
Barut-Girardello \cite{Barut} for the coherent states with the main aim of establishing the respective eigenfunctions 
$\{ \mathscr{F}_{\mu}(z;q) \}_{\mu \in \mathbb{C}}$ related to the lowering operator $\bop$. Such eigenfunctions were initially 
conceived as an infinite expansion of the Rogers-Szeg\"{o} functions whose coefficients satisfy a proper eigenvalue equation which leads
us to obtain the excitation probability distribution for the $q$-deformed coherent state. Besides, we evaluated the overlap 
probability and also discussed the completeness relation in this case. Hounkonnou and Ngompe Nkouankam \cite{Ngompe} recently showed an 
interesting study on the generalized hypergeometric coherent states, where a $(q,\nu)$-deformation was introduced in such a case. In that work, 
the authors employed basically some theoretical methods of quantum optics for investigating the quantum statistical properties as well as the 
Husimi distribution (and its corresponding phase distribution) of those particular coherent states. Such analysis can also be applied 
within the context here discussed, this fact being object of future investigations.

Finally, let us now briefly comment on the results obtained for the phase states. Basically, we have followed the Carruthers-Nieto approach
\cite{Carruthers} for the construction of the eigenstates related to the $q$-deformed cosine and sine operators, whose respective 
phase distributions have exhibited analogous behaviours but with distinct signatures (both the distributions are $\case{\pi}{2}$-dephased) 
in the $\gamma \varphi$ plane. Next, we have discussed two basic properties inherent to the eigenstates $\{ \mathscr{X}_{\gamma}(z;q) 
\}_{\gamma \in [0,\pi]}$ and $\{ \mathscr{Y}_{\gamma}(z;q) \}_{\gamma \in \lbk -\case{\pi}{2},\case{\pi}{2} \rbk}$ which reflect their
orthogonality and completeness (or resolution of unity) relations. In addition, we have applied our results in order to derive a set of 
closed-form expressions for certain mean values associated with the aforementioned cosine and sine operators via $q$-deformed coherent states,
which allow us to study the Robertson-Schr\"{o}dinger uncertainty relation. To complete this initial study, we have also derived a
symmetrical uncertainty relation (here involving the variances and covariances of the cosine, sine and nondeformed number operators) that
corroborates our previous conclusions on those coherent states: once the convergence criterion $0 \leq (1-q) | \mu |^{2} < 1$ be satisfied,
they are not minimum uncertainty states (excepting the vacuum state $| \mu | = 0$).

Although the mathematical significance of $q$ deformation is presently approached as being a parameter responsible for the distribution
width, its physical appeal is not quite clear. In fact, even the squeezing and/or nonlinear effects attributed to $q$ deserve to be 
investigated in details \cite{Bahri}. Furthermore, the difficulties in solving the intriguing problem related to the phase operator in 
quantum mechanics still remain the same \cite{Review}. In this sense, it is worth stressing that the compilation of results here presented 
not only corroborates and generalizes those obtained in \cite{Nelson,Carruthers}, but also represents a concatenated effort in joining two
promising research branches of mathematics and physics, namely the branches devoted to the study of $q$-special functions and certain
angular representations in quantum theory \cite{Ruzzi}.

In a more pragmatical sense, the previously discussed Rogers-Szeg\"{o} functions can be seen to be tailored for describing deformed physical
systems where the rotational degree of freedom has a central role. Indeed, some previous attempts of introducing rotational coherent states
have been put forth in the past whose aim was to treat the dynamics of two-dimensional deformed systems in molecular physics \cite{Reimers}.
In this connection, the use of the algebraic framework here developed in such studies of rigid deformed systems dynamics --- within the
context of von Neumann-Liouville formalism --- seems to be a promising perspective. Moreover, our results also seem to be quite suitable
to deal with the problem of quantum rings \cite{Lorke}, where a single electron can be trapped in a region whose topology is exactly that 
regarded in this work. In the meantime, it is worth stressing that there exists another way of future research which will be properly 
explored in due time.

%%%%%%%%%%%%%%%%%%%%%%%%%%%%%%%%%%%%%%%%%%%%%%%%%%%%%%%%%%%%%%%%%%%%%%%%%%%%%%%%%%%%%%%%%%%%%%%%%%%%%%%%%%%%%%%%%%%%%%%%%%%%%%%%%%%%%%%%%%
\ack
%%%%%%%%%%%%%%%%%%%%%%%%%%%%%%%%%%%%%%%%%%%%%%%%%%%%%%%%%%%%%%%%%%%%%%%%%%%%%%%%%%%%%%%%%%%%%%%%%%%%%%%%%%%%%%%%%%%%%%%%%%%%%%%%%%%%%%%%%%

The authors thank the anonymous referees for useful comments on an earlier version of this paper.  

%%%%%%%%%%%%%%%%%%%%%%%%%%%%%%%%%%%%%%%%%%%%%%%%%%%%%%%%%%%%%%%%%%%%%%%%%%%%%%%%%%%%%%%%%%%%%%%%%%%%%%%%%%%%%%%%%%%%%%%%%%%%%%%%%%%%%%%%%%

%%%%%%%%%%%%%%%%%%%%%%%%%%%%%%%%%%%%%%%%%%%%%%%%%%%%%%%%%%%%%%%%%%%%%%%%%%%%%%%%%%%%%%%%%%%%%%%%%%%%%%%%%%%%%%%%%%%%%%%%%%%%%%%%%%%%%%%%
\end{document}